\documentclass[prd,twocolumn,superscriptaddress,floatfix,amsmath,amssymb]{revtex4-1}
\usepackage[english]{babel}
\usepackage{graphicx}
\usepackage{dcolumn}
\usepackage{bm}
\usepackage{verbatim}
\usepackage{mathrsfs}
\usepackage{cancel}
\usepackage{color}

\newcommand{\integral}[3]{\int_{#2}^{#3} \text{d} #1}

\newcommand{\partialfrac}[2]{\frac{\partial #1}{\partial #2}}
\newcommand{\ket}[1]{\left| {#1} \right\rangle}
\newcommand{\bra}[1]{\left\langle {#1} \right|}

\newcommand{\proj}[2]{\left| {#1} \right\rangle\!\left\langle {#2} \right|}
\newcommand{\comm}[2]{\left[{#1},{#2}\right]}

\newcommand{\ii}{\mathrm{i}}

\newcommand{\tr}{\operatorname{Tr}}

\newcommand{\eq}[1]{\eqref{#1}}

\newcommand{\fourpoint}[8]{\left< \phi( x_{\mathcal{#1}}(#2)) \phi( x_{\mathcal{#3}}(#4)) \phi( x_{\mathcal{#5}}(#6)) \phi( x_{\mathcal{#7}}(#8)) \right>}
\newcommand{\intH}[1]{H_{I, \mathcal{#1}}}
\newcommand{\twopoint}[4]{\left< \phi( x_{\mathcal{#1}} (#2) ) \phi ( x_\mathcal{#3} (#4) ) \right>}
\newcommand{\mud}[1]{M_{\mathcal{#1}}}

\def\slashchar#1{\setbox0=\hbox{$#1$} 
\dimen0=\wd0 
\setbox1=\hbox{/} \dimen1=\wd1 
\ifdim\dimen0>\dimen1 
\rlap{\hbox to \dimen0{\hfil/\hfil}} 
#1 
\else 
\rlap{\hbox to \dimen1{\hfil$#1$\hfil}} 
/ 
\fi}

\begin{document}

\title{Quantum signalling in cavity QED} 

\author{Robert H. Jonsson}
\affiliation{Dept. Applied Math., University of Waterloo, 200 University
Av. W., Waterloo, Ontario, N2L 3G1, Canada}
\author{Eduardo Mart\'{i}n-Mart\'{i}nez}
\affiliation{Dept. Applied Math., University of Waterloo, 200 University
Av. W., Waterloo, Ontario, N2L 3G1, Canada}
\affiliation{Institute for Quantum Computing, University of Waterloo, Waterloo, Ontario, N2L 3G1, Canada}
\affiliation{Perimeter Institute for Theoretical Physics, 31 Caroline St N, Waterloo, ON, N2L 2Y5, Canada}
\author{Achim Kempf}
\affiliation{Dept. Applied Math., University of Waterloo, 200 University
Av. W., Waterloo, Ontario, N2L 3G1, Canada}
\affiliation{Institute for Quantum Computing, University of Waterloo, Waterloo, Ontario, N2L 3G1, Canada}
\affiliation{Perimeter Institute for Theoretical Physics, 31 Caroline St N, Waterloo, ON, N2L 2Y5, Canada}
\affiliation{Centre for Quantum Computation \& Communication Technology,
Dept. of Physics, University of Queensland, St Lucia QLD 4072, Australia}

\begin{abstract}


We consider quantum signalling between two-level quantum systems in a cavity, in the pertubative regime of the earliest possible arrival times of the signal.
 We present two main results: First we find that, perhaps surprisingly, the analogue of amplitude modulated signalling (Alice using her energy eigenstates $\ket{g}, \ket{e}$, as in the Fermi problem) is generally sub-optimal for communication. Namely, e.g., phase modulated signalling (Alice using, e.g., $\ket{+},\ket{-}$-states) overcomes the quantum noise already at a lower order in perturbation theory.  Second, we study the effect of mode truncations that are commonly used in cavity QED on the modelling of the communication between two-level atoms. We show that, on general grounds, namely for causality to be preserved, the UV cutoff must scale at least polynomially with the desired accuracy of the predictions.

\end{abstract}

\maketitle

\section{Introduction}

The interaction of electromagnetic radiation with matter is of fundamental as well as of practical importance. In practice, while a fundamental quantum field theoretic description is available, simplified models are often used. For example, the Jaynes-Cummings model, derived from the atomic dipolar coupling to the electromagnetic field under the rotating-wave and single mode approximations \cite{ScullyBook}, has proven to be adequate for the description of many quantum optics experiments. However, experimental advances keep pushing for refinements of the models used.

In particular, recent experimental techniques in quantum optics and superconducting circuits have allowed for ultrafast optical measurements, approaching time scales comparable with the inverse of the characteristic frequency of the atomic energy gap between the ground and the excited states. In this regime, two of the most commonly employed approximations break down: the rotating-wave approximation and the single mode approximation \cite{ScullyBook}.

Concretely, as has been pointed out in \cite{Fay,Fermicausality}, the naive use of a Jaynes-Cummings model with a single mode approximation (i.e., considering a system of detectors which interact only with a single mode of the quantum field in a cavity) shows inconsistencies at these scales, in the sense that it would allow superluminal signalling. The occurrence of this problem is plausible, considering that a single mode is a completely nonlocal degree of freedom. In principle, it takes all modes to describe arbitrarily localized interactions, and traveling waves, and therefore causal propagation. In \cite{UdWGauss} it was shown in a particular scenario using harmonic oscillators that causal behavior should be approximately recoverable when only a finite number of modes is taken into account. 

This leads to a question of practical importance. Namely, given an arbitrary level of desired accuracy, which finite minimum number of modes does one need to take into account in order to make the model consistent with causal communication? In order to address this question we will here study the fundamental quantum communication channel consisting of two localized quantum systems, such as atoms, that exchange field quanta in a cavity. 

On one hand, our approach will allow us to find the answer to the practical question posed: the number of modes that one needs to take into account needs to grow according to a certain power law with respect to the required accuracy with which communication is to be described as causal. 

On the other hand, the study of this quantum channel, this time without cutting off the number of modes, will here also lead us to a new basic observation in the regime of the earliest times when the signal can arrive. We will find that, in this regime, amplitude modulation is not optimal, i.e., that it can be improved upon through modulation in bases other than the eigenbasis of the sender's free Hamiltonian. 

With this unexpected result we extend a body of literature \cite{Hegerfeld,Ferm2,Ferm3,Fermicausality} that started with a question by Fermi \cite{Fermi}: How do quantum field theories guarantee causality, given that quantum field theoretic Feynman propagators necessarily \cite{Feynm} possess tails outside the light cone (which decay exponentially for massive fields and polynomially for massless fields)? Consider, for example, two atoms $\cal{A}$ (Alice) and $\cal{B}$ (Bob) at rest separated by a spatial distance $R$ at some common proper time $t=0$. $\cal{A}$ is prepared in an excited state while $\cal{B}$ is in the ground state and the electromagnetic field is in the vacuum state. Fermi's question is then ``Can the atom $\cal{A}$ decay to the ground state and induce an excitation of the atom $\cal{B}$ at a time $t<R/c$?''.  The answer has to be that the atom $\cal{B}$ has a non-zero probability of getting excited outside the light cone, but that this probability is completely independent of atom $\cal{A}$ \cite{Sab1,Sab2,Fermicausality}, so that no information is being carried over a spacelike distance. 

From a more modern perspective than Fermi's, namely that of quantum information theory, it is clear that nonlocal quantum correlations outside the light cone do not necessarily pose a problem. This is because the vacuum state is a spatially entangled state. In fact, it is known that two localized and spacelike separated quantum systems can be made entangled by merely letting them interact with the field vacuum state. The systems get entangled because they swap entanglement from the vacuum rather than by interacting through the exchange of real field quanta, see e.g., \cite{reznik,Hegerfeld, Sab1,Sabin2,Sab2,resin,mathieuachim1}. The impact of curvature was studied in \cite{ClicheKempfD}.   

The Fermi problem was  formulated entirely in a quantum information theoretic framework in \cite{mathieuachim1}, namely by defining and studying the quantum channel that maps the density matrix of Alice to the density matrix of Bob. There, it was shown that the classical and quantum capacities of the quantum channel from Alice to Bob exactly vanish outside the light cone to all orders in perturbation theory.  Here, we will use the quantum channel of \cite{mathieuachim1}, but with Alice and Bob in a cavity. 

First, we will study how many modes need to be taken into account in practical calculations so that causality is preserved to any desired accuracy.  We will find that arbitrary accuracy can be reached, in principle, (in spite of the Gibbs phenomenon) and we will find the scaling behavior of the number of modes needed. To obtain the most stringent bound, we will assume that the interaction is switched on suddenly and left on thereafter.  

Second, without cutting off the number of modes, we will consider the regime of the earliest times that the signal can arrive, where we will derive the unexpected result that Alice can do better than send her message through amplitude modulation. Namely, when she amplitude-codes her message by preparing a ground or an excited state, Bob only hears her as weakly as in 4th order of perturbation theory. If instead she ``phase-modulates" her signal by preparing instead for example $\vert +\rangle, \vert - \rangle$ states, Bob will receive a signal from Alice in second order of perturbation theory. Finally, we will also discuss the generalization from cavities to free space.

\section{Setting}\label{sec:setting}

Our aim is to analyze signal transmission from an atom $\mathcal{A}$ placed inside a cavity at some distance from a second atom $\mathcal{B}$ through the interaction with the field in the cavity.  As a first approximation to the problem let us consider that the atoms are point like (their characteristic size is much smaller than the smallest characteristic wavelength in the cavity). For realistic atoms in microwave and optical cavities this is a very good approximation as it is discussed, e.g., in \cite{Wavepackets}.

To model this situation we consider a pair of two-level quantum systems as our atoms $\mathcal{A}$ and $\mathcal{B}$ as it is commonplace in the literature \cite{reznik,Sab1,Sab2,PastFutPRL,Wavepackets,AasenPRL}. The interaction of an atom and the radiation inside a cavity can be  approximated (for atomic transitions with no exchange of angular momentum) by the Unruh-DeWitt Hamiltonian, \cite{Wavepackets}, which describes the interaction of a two-level system with a scalar field \cite{DeWitt}. The Hamiltonian is 
$H=H_0^{(\mathcal{D})}+H_0^{(\mathcal{F})}+H_I$, where $H_0^{(\mathcal{D})}$ and $H_0^{(\mathcal{F})}$ are the free Hamiltonians of  the two-level system and the field respectively, and $H_{I}$ is the interaction Hamiltonian:
\begin{equation}
H_{I}= \lambda \ \chi(\tau) \mu(\tau) \phi(x(\tau))\label{udw0},
\end{equation}
Here, $\lambda$ is the coupling strength and $\chi(\tau)$ is a switching function controlling the interaction time. The two-level system (which we will refer to as the 'detector' or 'atom') is coupled to the  massless scalar field $\phi (x)$  along its worldline (parametrized in terms of its proper time $\tau$) $x(\tau)$ through its monopole moment $\mu(\tau)$.

In the case where we have two inertial detectors $\mathcal{A, B}$ placed at constant positions $x_\mathcal{A}$ and $x_\mathcal{B}$ in a stationary cavity of length $L$,  with the usual mode expansion for the field, the interaction Hamiltonian can be written as 
\begin{align}\label{hamilto}
H_{I} &=  \sum_{\mathcal{D}=\mathcal{A,B}}\lambda_\mathcal{D}\, \chi_\mathcal{D}(t) \mu_\mathcal{D}(t) \,  \nonumber\\
&\qquad \quad \times\sum_{j=1}^{\infty} (a_{j}^{\dagger}e^{i \omega_{j} t}+a_{j} e^{-i \omega_{j} t})\frac{\sin{k_j x_\mathcal{D}}}{\sqrt{\omega_jL}}
\end{align}
in the interaction picture. 
Notice, that the proper time of the detector now coincides with the time coordinate $t$ of the cavity's frame in which the field quantization was carried out.
Now, the monopole moment of the two-level detectors takes the usual form
\[\mu_\mathcal{D} (t)=\sigma_\mathcal{D}^{+} \, e^{i \Omega_\mathcal{D} t}+\sigma_\mathcal{D}^{-} \, e^{-i \Omega_\mathcal{D} t},\]
where $\Omega_\mathcal{D}$ is the energy difference (or gap) between the ground and excited state of the detector $\mathcal{D}$, and   $\sigma_\mathcal{D}^+=\ket{e_\mathcal{D}}\bra{g_\mathcal{D}}$ and $\sigma_\mathcal{D}^- = \ket{g_\mathcal{D}}\bra{e_\mathcal{D}}$. 

As can be  seen easily, the Jaynes-Cummings model that describes the radiation-matter interaction commonly used in quantum optics is essentially what is obtained by taking a single-mode approximation in $H_I$ \cite{ScullyBook}.

The communication between Alice and Bob is modelled as follows: Initially, the switching functions are chosen to vanish. So the detectors are not coupled to the field and Alice and Bob can prepare them for the interaction. To encode her message, Alice is free to prepare her detector in any state $\rho_{\mathcal{A},0}$ she chooses. Bob prepares his detector in the ground state and the field is assumed to start out in the vacuum. Hence the initial state of the system is
\begin{equation}\label{eq:instate}
\rho_{0} = \rho_{\mathcal{A}, 0} \otimes \ket{g_\mathcal{B}} \bra{g_\mathcal{B}} \otimes \ket0 \bra0.
\end{equation}
The detectors are then coupled to the field within a time interval $t\in (0,T)$. In this interval the initial state evolves under the action of $H_\text{int}$ in general into an entangled output state $\rho_T$.

As we are interested in the communication between Alice and Bob, we are interested in how the output state of Bob's detector is dependent on Alice's choice of $\rho_{\mathcal{A},0}$.
This gives rise to a quantum channel map from Alice's input density matrix to Bob's output density matrix
\begin{equation}\label{eq:channel}
\xi: \rho_{\mathcal{A}, 0} \mapsto \rho_{\mathcal{B}, T}=\tr_{\mathcal{A}} \tr_\mathcal{F} \left( \rho_T \right)
\end{equation}
where we obtain Bob's output state by taking the partial trace over Alice's and the field's Hilbert space.

The time evolution operator under this Hamiltonian from a time $t=0$ to a time $t=T$ is given by the following expansion:
\begin{multline}\label{eq:pert}
U(T,0) = \openone\!\underbrace{-\ii\int_{0}^{T}\!\!\!\!\!d t_1 H_{I}(t_1)}_{U^{(1)}} \\
\underbrace{ - \int_{0}^{T}\!\!\!\!\!dt_1 \!\!\int_{0}^{t_1}\!\!\!\!\!dt_2\,    H_{I}(t_1) H_{I}(t_{2})}_{U^{(2)}}+\hdots
\end{multline}
Under the realistic assumption that the coupling strength is small enough, we can neglect higher orders in \eqref{eq:pert}. If we denote by $\rho_0$ the initial density matrix of the field-detector system we get that after a time $T$, 
\begin{equation}
\rho_T=[\openone+U^{(1)}+U^{(2)}+\mathcal{O}(\lambda^3)]\rho_0[\openone+U^{(1)}+U^{(2)}+\mathcal{O}(\lambda^3)]^{\dagger}
\end{equation}
This is, $\rho_T=\rho_0+\rho_T^{(1)}+\rho_T^{(2)}+\rho_T^{(3)}+\rho_T^{(4)}+\mathcal{O}(\lambda^5)$, where
\begin{align}
\label{eq:rho1}\rho_T^{(1)}&=U^{(1)}\rho_0+\rho_0{U^{(1)}}^\dagger \sim\mathcal{O}(\lambda)\\
\label{eq:rho2}\rho_T^{(2)}&=U^{(1)}\rho_0{U^{(1)}}^\dagger+U^{(2)}\rho_0+\rho_0{U^{(2)}}^\dagger \sim\mathcal{O}(\lambda^2)\\
& \dots \nonumber \\
\label{eq:rhogen}\rho_T^{(n)}&=\sum_{k=0}^n \, U^{(n-k)} \, \rho_0 \, {U^{(k)}}^\dagger \sim\mathcal{O}(\lambda^n)
\end{align}
and $U^{(0)}=\openone$ is understood. 
The symbol $\mathcal{O}(\lambda^n)$ stands for the combined powers of the two coupling constants, i.e., $\mathcal{O}(\lambda_\mathcal{A}^i \lambda_\mathcal{B}^j)\sim \mathcal{O}(\lambda^{i+j})$.

In \cite{mathieuachim1}, the time evolution was formulated in terms of commutators between 
$H_I$ and $\rho_0$. We here choose to use the Dyson expansion of $U(t)$ directly instead, because it facilitates the intuitive  interpretation of the different perturbative processes  and leads to an integral structure that is easier to evaluate numerically.

Note that  all the  perturbative corrections from \eqref{eq:rhogen} to the final density matrix $\rho_T$ are traceless:
\begin{equation}\label{eq:traceless}
\tr \rho_T^{(n)}=0.
\end{equation}
Therefore, independent of up to which order $\mathcal{O}(\lambda^n)$ in the coupling constant the corrections are taken into account, the trace of the final state is always preserved, 
\begin{equation}
\tr \rho_T =1,
\end{equation}
and no normalization constant is necessary in front of $\rho_T$ at any given order in perturbation theory if all the terms of a given order are consistently taken into account. To see this, let us verify that the derivative of the left hand side of \eqref{eq:traceless} with respect to the switching time $T$ vanishes:
\begin{equation}\label{proof1}
\partialfrac{}T \tr \rho_T^{(n)}=0,\; \forall\, T\geq 0.
\end{equation}
As all the $\rho_T^{(n)}$ are identically zero for $T=0$, their trace also vanishes for $T=0$. Hence, if \eqref{proof1} is true then $\tr \rho_T^{(n)}$ vanishes for all $T$. To evaluate \eqref{proof1}  we differentiate \eqref{eq:rhogen}:
\begin{multline}\label{eq:trrhoderiv}
\partialfrac{}T\tr \rho_T^{(n)} =\tr \left[ \left(-\ii\, H_I(T) U^{(n-1)}\right) \rho_0 \right.  \\ + \sum_{k=1}^{n-1} \left(-\ii\, H_I(T) U^{(n-k-1)}\right) \rho_0 \, {U^{(k)}}^\dagger \\ +\sum_{k=1}^{n-1} U^{(n-k)} \,  \rho_0 \left(\ii\, {U^{(k-1)}}^\dagger  H_I(T) \right) \\ \left. + \rho_0\left(\ii\, {U^{(n-1)}}^\dagger  H_I(T) \right) \right]
\end{multline}
where we used $\partialfrac{}{T} U^{(n)}= -\ii\, H_I(T) U^{(n-1)}$, which follows from \eqref{eq:pert}. Using the cyclic property of the trace, we can rewrite \eqref{eq:trrhoderiv} so as to have $H_I(T)$ stand, e.g., in front of every term in the sum. Then we see that the terms form pairs that exactly cancel each other, so \eqref{eq:trrhoderiv} vanishes. This shows that $\tr \rho_T^{(n)}$ is indepent of $T$ and vanishes for all $T$. 

Of course, independently of that, the Dyson expansion is not unitary order by order, but instead it is unitary up to the power of the perturbative parameter of the first ignored term in the perturbative expansion. Additionally, it is well-known that for very long times we can leave of the perturbative regime (where the Dyson expansion is valid) \cite{Wiilliam}. However, in the current paper we are concerned with timescales smaller or of the order of the light-crossing time of the cavity, and for all the values considered here, the Dyson expansions will be unitary up to at least the 4th (or 6th) power of the perturbative parameter.

Now to obtain Bob's output density matrix $\rho_{B, T}$, as defined in \eq{eq:channel}, we trace out the field and Alice's detector from $\rho_T$. However, all $\rho_T^{(n)}$, with $n$ odd, do not contain diagonal matrix elements in the field components, hence they drop out when the partial trace over the field is taken. This is because the field starts out in the vaccuum state, hence the partial trace over the field can be expressed as a vaccuum $n$-point function of the field. These $n$-point functions are identically zero for odd numbers of field operators. So the contributions to $\rho_{B, T}$ are all of even power in the coupling constant.
\begin{equation} \label{eq:rhobout}
\rho_{B, T} =\rho_{B,0} + \tr_{{A},\mathcal{F}} \rho_T^{(2)}+ \tr_{\mathcal{A},\mathcal{F}} \rho_T^{(4)} + \mathcal{O}(\lambda^6)
\end{equation}

The dependence of Bob's output density matrix on the elements of Alice's input density matrix is captured by the quantum channel $\xi$ from \eq{eq:channel}. 
\begin{align}
\xi \left[ \rho_{A, 0} \right] = \rho_{B, T}
\end{align}•
Denoting Alice's initial density matrix as
\begin{align}\label{eq:defnrhoA}
\rho_{A,0} &= \theta \ket e \bra e + \gamma \ket e \bra g + \gamma^* \ket g \bra e + \beta \ket g \bra g\nonumber\\
&= \begin{pmatrix} \theta & \gamma \\ \gamma^* & \beta \end{pmatrix},
\end{align}
it's general structure   is given by \cite{mathieuachim1}
\begin{align}\label{eq:chanstruct}
\xi \left[ \begin{pmatrix}
\theta & \gamma \\ \gamma^* & \beta \end{pmatrix} \right] 
&= \begin{pmatrix}
P & 0 \\ 0 & 1-P
\end{pmatrix} \nonumber\\
&\qquad + \begin{pmatrix} \theta A +\beta B & \gamma C + \gamma^* D^* \\ \gamma^* C^* +\gamma D & -\theta A - \beta B
\end{pmatrix}.
\end{align}

 The term $P$ accounts for the  noise observed by Bob and is independent of the presence of Alice's detector. Indeed, it is not affected by the elements of the density matrix of $\mathcal{A}$ in \eqref{eq:chanstruct}.
The terms that account for the influence of detector $\mathcal{A}$ on $\mathcal{B}$ are those labeled $A,B,C$ and $D$.
$A, B$ and $P$ are real, while $C$ and $D$ are complex. They depend on the parameters of the detectors  and cavity and the switching function. Their lowest order contributions are:
\begin{align}
\label{eq:P}P&= \lambda_\mathcal{B}^2 \,  P_2 + \lambda_\mathcal{B}^4 \, P_4 +\mathcal{O}(\lambda_\mathcal{B}^6)\displaybreak[0]\\
\label{eq:A}A&=\lambda_\mathcal{A}^2 \lambda_\mathcal{B}^2 \, A_4+\mathcal{O}(\lambda^6)\displaybreak[0]\\
\label{eq:B} B&=\lambda_\mathcal{A}^2 \lambda_\mathcal{B}^2 \, B_4+\mathcal{O}(\lambda^6)\displaybreak[0]\\
\label{eq:C}C&= \lambda_\mathcal{A} \lambda_\mathcal{B} \, C_2 + \mathcal{O}(\lambda^4)\displaybreak[0]\\
\label{eq:D} D&=\lambda_\mathcal{A} \lambda_\mathcal{B} \, D_2 + \mathcal{O}(\lambda^4)
\end{align}
Here again the symbol $\mathcal{O}(\lambda^n)$ stands for the combined powers of the two coupling constants. The expressions for $C_2,D_2,A_4,B_4,P_2$ and $P_4$ are rather complex and are given in appendix \ref{AppA} where some interesting points about their mathematical form are also discussed.

We can understand both the general structure of the channel and the form of the individual terms by discussing how they originate from \eq{eq:rhobout}.

Every term $U^{(k)}$ in the expansion of the time evolution operator in \eqref{eq:pert} can be expanded into $2^k$ summands, by using that the interaction Hamiltonian \eqref{hamilto} is the sum $H_I = H_{\mathcal{A},I} +H_{\mathcal{B},I}$ of the interaction Hamiltonian for each of the detectors. Accordingly, each $\rho_T^{(k)}$ can be written as a sum of terms sorted by their orders in the coupling constants $\lambda_\mathcal{A}$ and $\lambda_\mathcal{B}$.


 In this fashion, the lowest order contribution to $\rho_{B,T}$, which reads
\begin{eqnarray}\label{eq:rbt2}
\rho_{\mathcal{B},T}^{(2)} &=&  \tr_{\mathcal{A},\mathcal{F}} \, \rho_T^{(2)} \nonumber\\ &=&\tr_{\mathcal{A},\mathcal{F}} \left[ U^{(1)}\rho_0{U^{(1)}}^\dagger+U^{(2)}\rho_0+\rho_0{U^{(2)}}^\dagger \right],
\end{eqnarray}
contains terms of order $\mathcal{O}(\lambda_\mathcal{A}^2), \, \mathcal{O}(\lambda_\mathcal{B}^2)$ and $\mathcal{O}(\lambda_\mathcal{A} \lambda_\mathcal{B})$.

The terms of order $\mathcal{O}(\lambda_\mathcal{A}^2)$ do not contribute to $\rho_{B,T}$  because they cancel out when the partial trace over detector $\mathcal{A}$ is taken.  This holds true for all terms that do not contain any power of $\lambda_\mathcal{B}$, hence no terms of order $\mathcal{O}(\lambda_\mathcal{A}^n)$ contribute to $\rho_{B,T}$.

The terms of order $\mathcal{O}(\lambda_\mathcal{B}^2)$ contribute to either the upper or to the lower diagonal element of $\rho_{B,T}$. The contribution of this kind originating from $U^{(1)} \rho_0 {U^{(1)}}^\dagger$ is proportional to $\proj{e_\mathcal{B}}{e_\mathcal{B}}$, while the $\mathcal{O}(\lambda_\mathcal{B}^2)$ contribution  from $U^{(2)}\rho_0+\rho_0{U^{(2)}}^\dagger$ leads to terms that are proportional to $\proj{g_\mathcal{B}}{g_\mathcal{B}}$.  Although they come with different integral structures, the coefficients of these matrix elements are equal up to an overall sign. They both constitute $P_2$, the lowest order contribution to $P$.

It is important to remark that,  as mentioned above, $P$ is nothing but the excitation probability of the single detector in the quantum vacuum. This quantum noise term is  independent of the presence of the second detector and it contains only  terms of order $\mathcal{O}(\lambda_\mathcal{B}^n)$.

Any terms that describe an interaction between the two detectors have to contain powers of both coupling constants, i.e., they are $\mathcal{O}(\lambda_\mathcal{A}^i \lambda_\mathcal{B}^j)$. 

The terms of order $\mathcal{O}(\lambda_\mathcal{A}^i \lambda_\mathcal{B}^j)$, with $i$ and $j$ odd, always appear multiplied by $\gamma$ or $\gamma^*$ (off-diagonal elements of A's initial state) and  $\proj{e_\mathcal{B}}{g_\mathcal{B}}$ or $\proj{g_\mathcal{B}}{e_\mathcal{B}}$ (off-diagonal elements of B's final state), so they contribute to the factors $C$ or $D$, which couple the off-diagonal elements of $\rho_{\mathcal{A},0}$ and $\rho_{B,T}$ as we see in the general structure of the quantum channel \eqref{eq:channel}.

This means that the terms of order  $\mathcal{O}(\lambda_\mathcal{A} \lambda_\mathcal{B} )$ from \eqref{eq:rbt2} are the lowest order terms that account for any signalling from A to B if the initial state of Alice's detector is such that $\gamma\neq0$.

As all the terms contributing to $\rho_{B,T}$ are of even (combined) powers in the coupling constant, the only other class of terms contributing to the channel are of order $\mathcal{O}(\lambda_\mathcal{A}^i \lambda_\mathcal{B}^j)$, with both $i$ and $j$ even. These terms couple the diagonal terms of both density matrices and hence contribute to the factors $A$ and $B$ in \eqref{eq:channel}. The lowest order terms in this class are of order $\mathcal{O}(\lambda_\mathcal{A}^2 \lambda_\mathcal{B}^2)$ given in \eqref{eq:A} and \eqref{eq:B}.

These observations have implications for the study of the causal behaviour of the setting. As mentioned above,  the leading contribution in order to study causality  depend on the initial state of the atoms and will, in general, be of order $\mathcal{O}(\lambda^2)$.  However, in the particular case where the initial density matrix of $\mathcal{A}$ is diagonal as, for instance, in the Fermi problem, the causal contributions are only of order $\mathcal{O}(\lambda^4)$ in the coupling strength and can hence be overpowered  by the noise term $P$, which is of second order. We will discuss this  more deeply below.


\section{Study of the quantum channel}

In the following we study the quantum channel \eqref{eq:chanstruct} between two detectors inside a cavity in different communication settings by  evaluating the leading order contributions to the factors $P, A, B, C, D$.

 As discussed previously, we know that for the model to be causal we need to consider the infinite number of field modes in the cavity. This is both unpractical and physically weakly founded since realistic cavities are not good cavities for the whole frequency spectrum. That seems to imply that the usual light-matter interaction model in cavities  violates causality and might therefore fail to describe the real light-matter interaction in realistic  cavities, above all if we are concerned with the transmission of information.  
 
Nevertheless, we know from quantum optics that models of cavities with a finite (and  often small) number of levels provide a very good approximation to the experimental phenomenology \cite{ScullyBook}. How good this approximation is depends on the timescales considered in the experiment, as discussed in the introduction.

Here we are interested in how the magnitude of the imposed UV-cutoff, i.e., the number of field modes that are taken into account, affects the accuracy with which causality is respected by the model.  In other words, we will study the magnitude of faster-than-light error terms as a function of the UV-cutoff.

As a first setting, we will consider what we will call the Fermi problem scenario, i.e., we study signalling from detector $\mathcal{A}$ to detector $\mathcal{B}$ under the condition that the initial state of the first detector is either the ground or the excited state. Although it appears to be a very natural choice to use the energy eigenstates for signalling, we know from the previous section that these signalling terms are suppressed by two orders in the coupling constant: here, the effect on Bob's detector is only of order $\mathcal{O}(\lambda^4)$, whereas the effect is of order $\mathcal{O}(\lambda^2)$ for any other set of pure input states.

To illustrate this, we will consider a second scenario where detector $\mathcal{A}$ initially is prepared in either the state
\begin{equation}
\ket{+}=\frac{1}{\sqrt{2}}\left(\ket{g} +\ket{e}\right)
\end{equation}
or
\begin{equation}
\ket{-}=\frac{1}{\sqrt{2}}\left(\ket{g} -\ket{e}\right).
\end{equation}

In what follows we use natural units $\hbar = c =1$ and consider a massless Klein-Gordon field inside a one-dimensional cavity. Hence $\omega_j = k_j = j \pi / L$ in \eqref{hamilto}.
Also we choose both detectors to be resonant with a field mode, so $\Omega_\mathcal{A} = \Omega_\mathcal{B} = \omega_n$ for some given resonance mode number $n$.

The detectors are switched on and off sharply at $t=0$ and $t=T$ respectively, i.e., the switching function is defined to be $\chi(t) = 1$ for $t \in (0,T)$ and vanish at all other times. Under  these conditions the perturbative terms in \eqref{eq:P}, \eqref{eq:A}, \eqref{eq:B}, \eqref{eq:C}, \eqref{eq:D} are analytically integrable (although very involved to obtain).  Even though different switching protocols could be considered, to have detector $\mathcal{B}$ switched in parallel with the first detector is the most conservative setting in order to detect any error terms that would propagate signals from $\mathcal{A}$ to $\mathcal{B}$ outside the lightcone.

 \begin{figure}[htb] 
\includegraphics[width=0.48\textwidth]{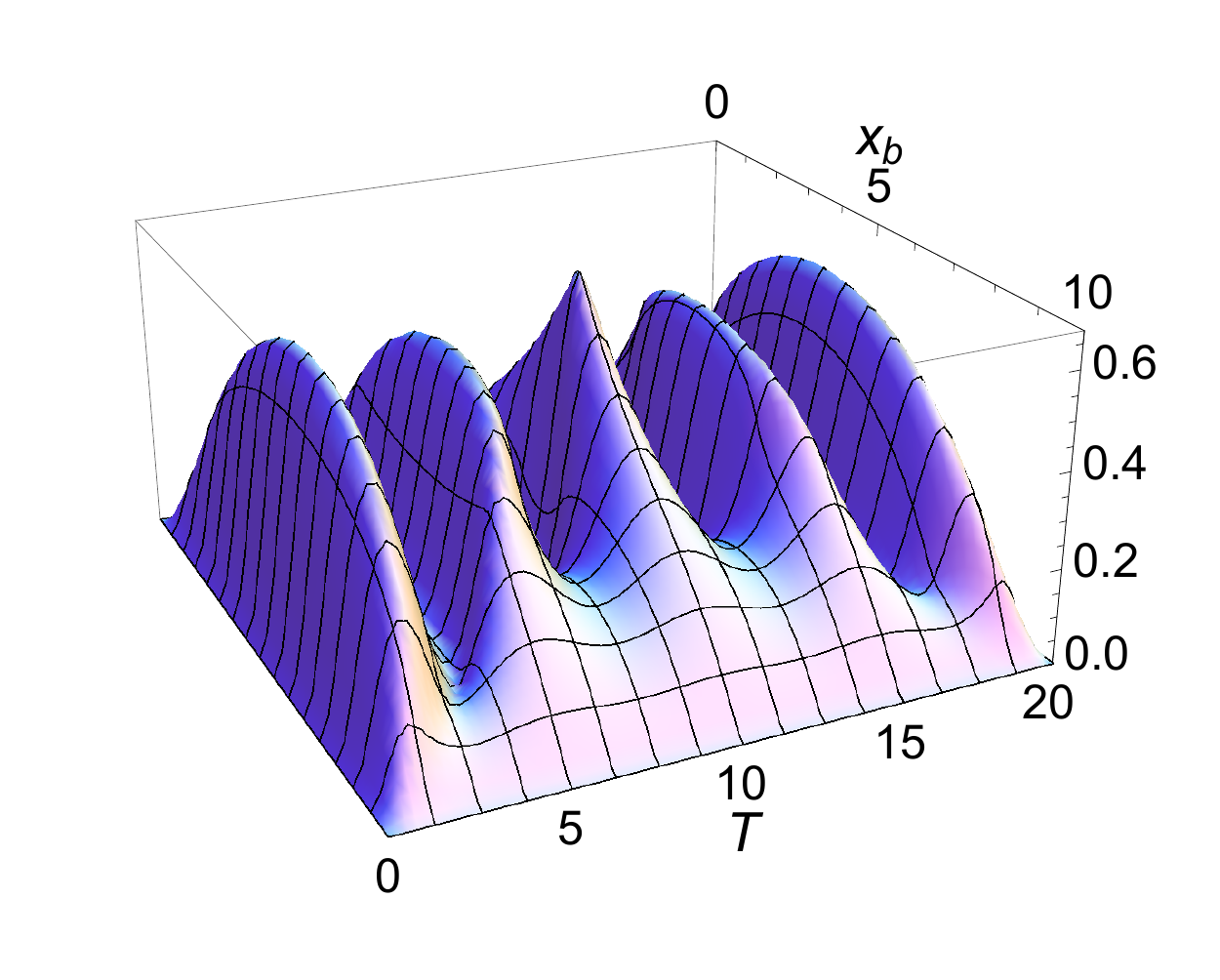}
  \caption{(Color online) Leading order contribution $P_2$ to the single detector excitation probability \eqref{eq:P} for a detector at $x_b$ in a cavity of length $L=10$. The detector is resonant to the fourth  field mode ($n=4$). The contribution $P_2$ is periodic with a periodicity of $T_\text{per} = 2L$. The number of valleys in every period is equal to $n$. 
 For the calculation a cutoff of $N_C=100$ modes was used. (All plotted quantities are dimensionless.)
}
  \label{fig:P2}
\end{figure}

Before we study the influence of detector $\mathcal{A}$ on the final state of $\mathcal{B}$ we review the contribution to $\rho_{\mathcal{B},T}$ in \eqref{eq:chanstruct} which is independent of the presence of detector $\mathcal{A}$. This is the term $P$ in \eqref{eq:P} which captures the probability of the single detector $\mathcal{B}$ to get excited on its own due to the switching, i.e., which captures its vacuum noise.

 Figure \ref{fig:P2} shows the lowest order contribution $P_2$ to the single detector excitation probability which is mostly induced from the vacuum due to the sudden switching. It is non-negative and periodic with a periodicity of $T_\text{per}=2 L$, which is twice the cavity crossing time.
When the detector is tuned resonant to a field mode with an even mode number $n$ (as in the figure), the term $P_2$ peaks at the light-crossing time of the cavity for a detector positioned at the middle of the cavity. If $n$ is odd $P_2$ vanishes here. The number of valleys per period along the $T$-axis is equal to the resonance mode number $n$. The contribution of the non-resonant modes is what makes the probability non-zero in these valleys.

In figure \ref{fig:A4B4P4}, one example of the second order contribution $P_4$ is plotted.
$P_4$ captivates the probability of a single detector to get excited and deexcited again during the interaction interval. Hence it gives a non-positive correction to the single detector excitation probability.

All other contributions to $\rho_{\mathcal{B},T}$ depend on the initial state of  Alice's detector. 

As pointed out above, in the Fermi problem the contribution of $\mathcal{A}$ to the state of $\mathcal{B}$ appears only in the diagonal elements of Bob's density matrix. Hence, as discussed earlier, in the Fermi problem the signalling terms which are of order $\mathcal{O}(\lambda^4)$ compete directly with the single detector excitation probability $P \sim \mathcal{O}(\lambda^2)$ which might on the one hand masks any effects of causality violations in the probability of excitation of Bob and on the other hand hinders the ability of Alice to signal Bob.

In the second example that we will consider by measuring in the $\{\ket{+}, \ket{-}\}$ basis Bob can detect the $\mathcal{O}(\lambda^2)$ effect of Alice's input  without any influence of $P$ at all on the measurement outcomes. This effect will be relevant if we want to use the field to transmit information from Alice to Bob.

\subsection{Signalling in the Fermi problem} 
We can analyze the Fermi problem, i.e., the question of how the excitation probability of detector $\mathcal{B}$, which starts out in the ground state at $t=0$, is affected by the presence of the other detector starting out in its excited state. From \eqref{eq:chanstruct} we see that the detector $\mathcal{B}$ ends up in the state
\begin{equation}
\xi \left( \proj{e_\mathcal{A}}{e_\mathcal{A}} \right) = \begin{pmatrix}
P+A & 0 \\ 0 & 1-P-A
\end{pmatrix}.
\end{equation}
So the factor $A$ describes the probability for the detector $\mathcal{B}$ to become excited due to the presence of the initially excited detector $\mathcal{A}$.
If we compare this output to the case where the detector $\mathcal{A}$ is initially prepared in its ground state,
\begin{equation}
\xi \left( \proj{g_\mathcal{A}}{g_\mathcal{A}} \right) = \begin{pmatrix}
P+B & 0 \\ 0 & 1-P-B
\end{pmatrix},
\end{equation}
we see that $B$ describes the contribution to the probability of finding detector $\mathcal{B}$ excited after the interaction due to the presence of the other detector starting out in the ground state.

 \begin{figure}[htb] 
\includegraphics[width=0.48\textwidth]{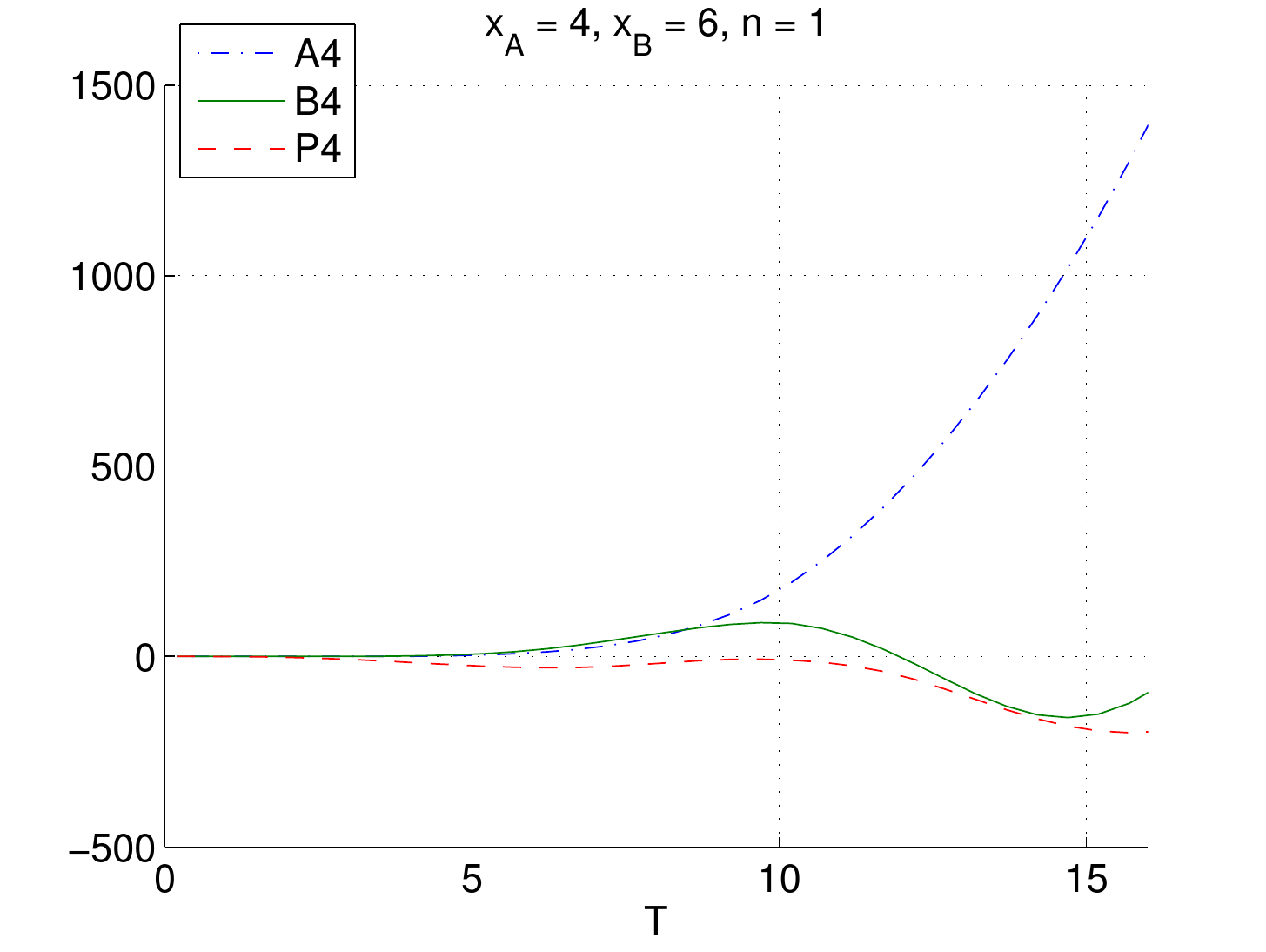}
  \caption{(Color online) Numerical values of the $\mathcal{O}(\lambda^4)$ coefficients defined in \eqref{eq:A},\eqref{eq:B} and \eqref{eq:P} for the quantum channel in the Fermi problem. The length of the cavity is $L=10$ and the distance between the two detectors is $|x_\mathcal{A} - x_\mathcal{B} | =2$. (All plotted quantities are dimensionless.)}
  \label{fig:A4B4P4}
\end{figure}

As we have seen in \eqref{eq:A} and \eqref{eq:B}, the factors $A$ and $B$ are of order $\mathcal{O}(\lambda^4)$. I.e., if Alice wants to send a message, or just a single bit, to Bob and tries to encode it by either preparing her detector in the ground or excited state initially, she only influences Bob's final measurement result at fourth order $\mathcal{O}(\lambda^4)$ in the coupling constant.

In figure \ref{fig:A4B4P4}, one example of the lowest order contributions to $A$ and $B$ is plotted. The general behaviour is that $A_4$ is non-negative and grows faster with the switching length $T$ than the other contributions of order $\mathcal{O}(\lambda^4)$, whereas $B_4$ is oscillating.

We also see that $A_4$ and $B_4$ vanish outside the lightcone, i.e., for switching times $T<|x_\mathcal{A}-x_\mathcal{B}|$ smaller than the distance between the two detectors. Of course this is necessary to prevent superluminal signalling: If $A$ or $B$ were not to vanish for $T<|x_\mathcal{A}-x_\mathcal{B}|$ then the state of detector $\mathcal{B}$ at time $t=T$ would be influenced by the the state of detector $\mathcal{A}$ at $t=0$, and thus retrieve information about the initial state of  $\mathcal{A}$, although no light signal could have reached $\mathcal{B}$ within this time.

It was shown in \cite{mathieuachim1} that the quantum channel $\xi$ is causal in the continuum scenario at leading order in perturbation theory. All the factors $A,B,C$ and $D$ in \eqref{eq:chanstruct} vanish outside the lightcone, which relies on the property of the field commutator to vanish for spacelike separations.

This also holds in the cavity if all (infinite) field modes are taken into account. However, when a UV-cutoff  is introduced such that only a finite number of modes $N_C$ are taken into account, the commutator does not vanish outside the lightcone any longer. Hence a model with only a finite number of field mode also predicts that superluminal signalling between two detectors is possible for certain settings. In the following we want to investigate how these acausalities depend on the number of modes $N_C$ and how the model of light-matter interaction behaves more and more causally with increasing cutoffs to a point where the predicted acausal behaviour would be undetectable in practice.

 \begin{figure}[htb] 
\includegraphics[width=0.48\textwidth]{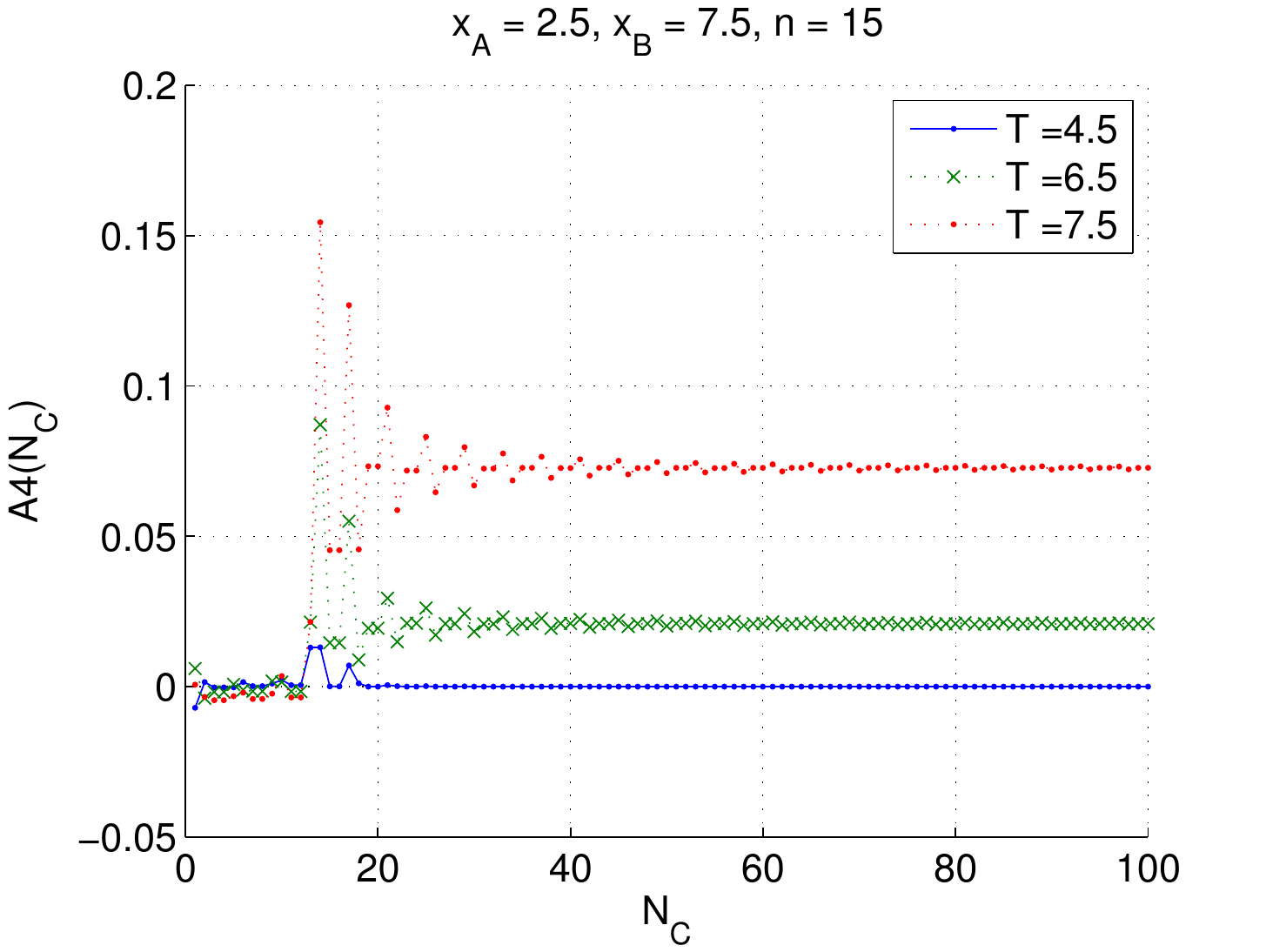}
  \caption{(Color online) Numerical values of the lowest order contribution $A_4$ from \eqref{eq:A} to the signalling term in the Fermi problem for two detectors separated by $|x_A-x_B|=5$ for different switching times $T$, depending on the number of modes below the cutoff $N_C$. The biggest contribution are acquired around the resonance mode number $n=15$. For $N_C > n$ the results oscillate around a limiting value which is approached for higher cutoffs.  For the lowest switching time the detectors are spacelike separated during the interaciton with the field, hence no signalling is possible. (All plotted quantities are dimensionless.)}
  \label{fig:A4ofNC}
\end{figure}

Figure \ref{fig:A4ofNC} illustrates how the result obtained for the coefficient in the channel, in this case of $A$, improve with the number of modes $N_C$ taken into account. As expected the main contribution to the coefficient originate from the mode to which the detectors are resonant. For cutoffs $N_C>n$ larger than the resonance mode the results begin to converge toward a limit in an oscillating manner. In figure \ref{fig:A4ofNC} this limit is positive because $T$ is chosen to be larger than the distance between the operators so that signalling is possible. If we have $T\leq |x_\mathcal{A}-x_\mathcal{B}|$ then the results for $A_4$ converge to zero.

\subsubsection*{Non-causal error terms decay with a power law in UV cutoff}

 \begin{figure}[htb] 
\includegraphics[width=0.48\textwidth]{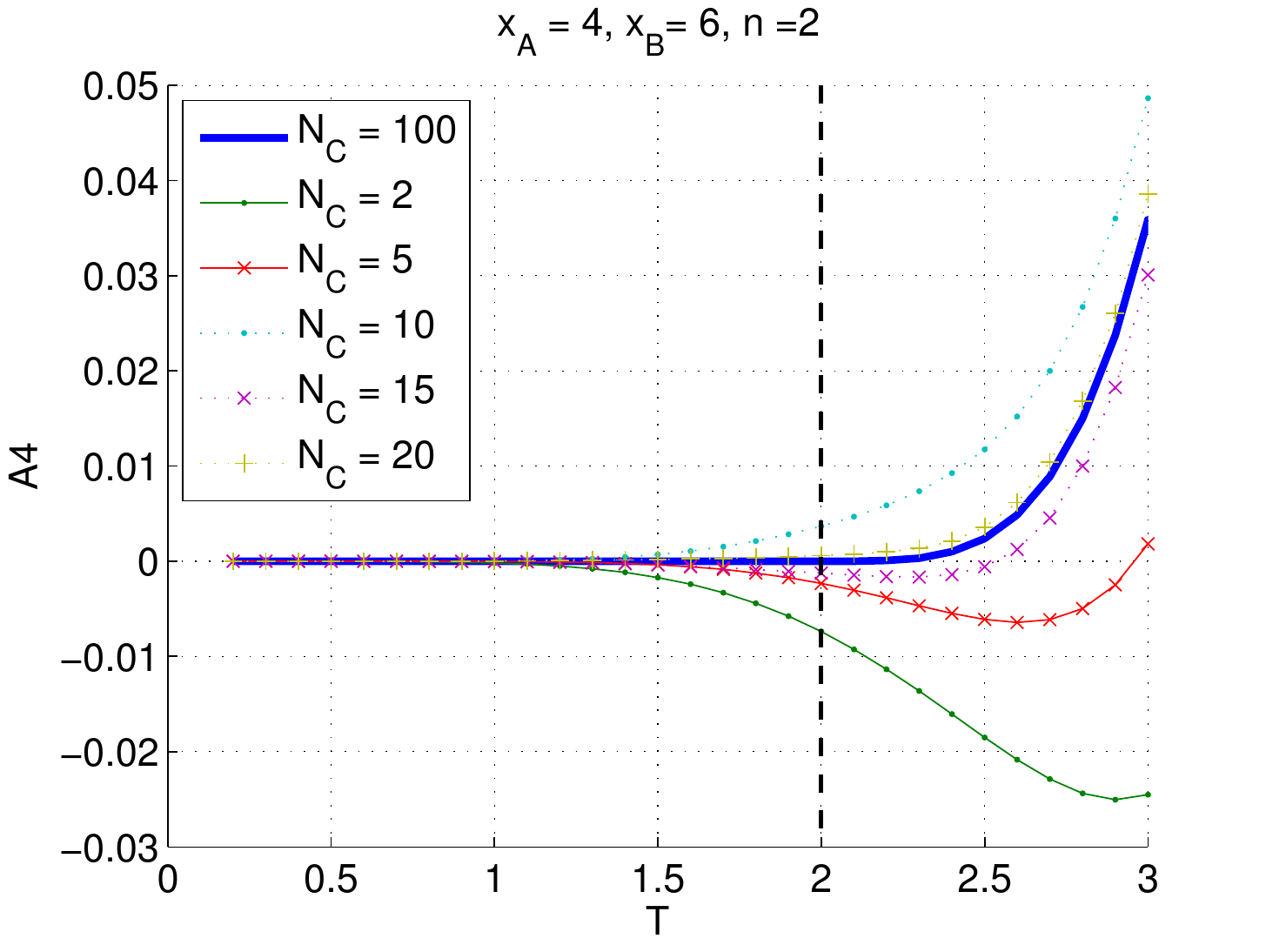}
  \caption{(Color online) The signalling term $A_4$ from \eqref{eq:A} in the Fermi problem for two detectors at a distance of $|x_\mathcal{A} - x_\mathcal{B} | =2$ for different switching times $T$ for increasing cutoffs at $N_C$. The dashed line indicates the lightcone. In general the values of $A_4$ inside the lightcone grow towards the lightcone. Hence to check the level of causality violation for a specific cutoff the value on the lightcone, i.e.,  for a switching time $T= |x_\mathcal{A} - x_\mathcal{B}|$ is relevant. (All plotted quantities are dimensionless.)}
  \label{fig:A4ofT}
\end{figure}

\begin{figure}[htb]
\includegraphics[width=0.48\textwidth]{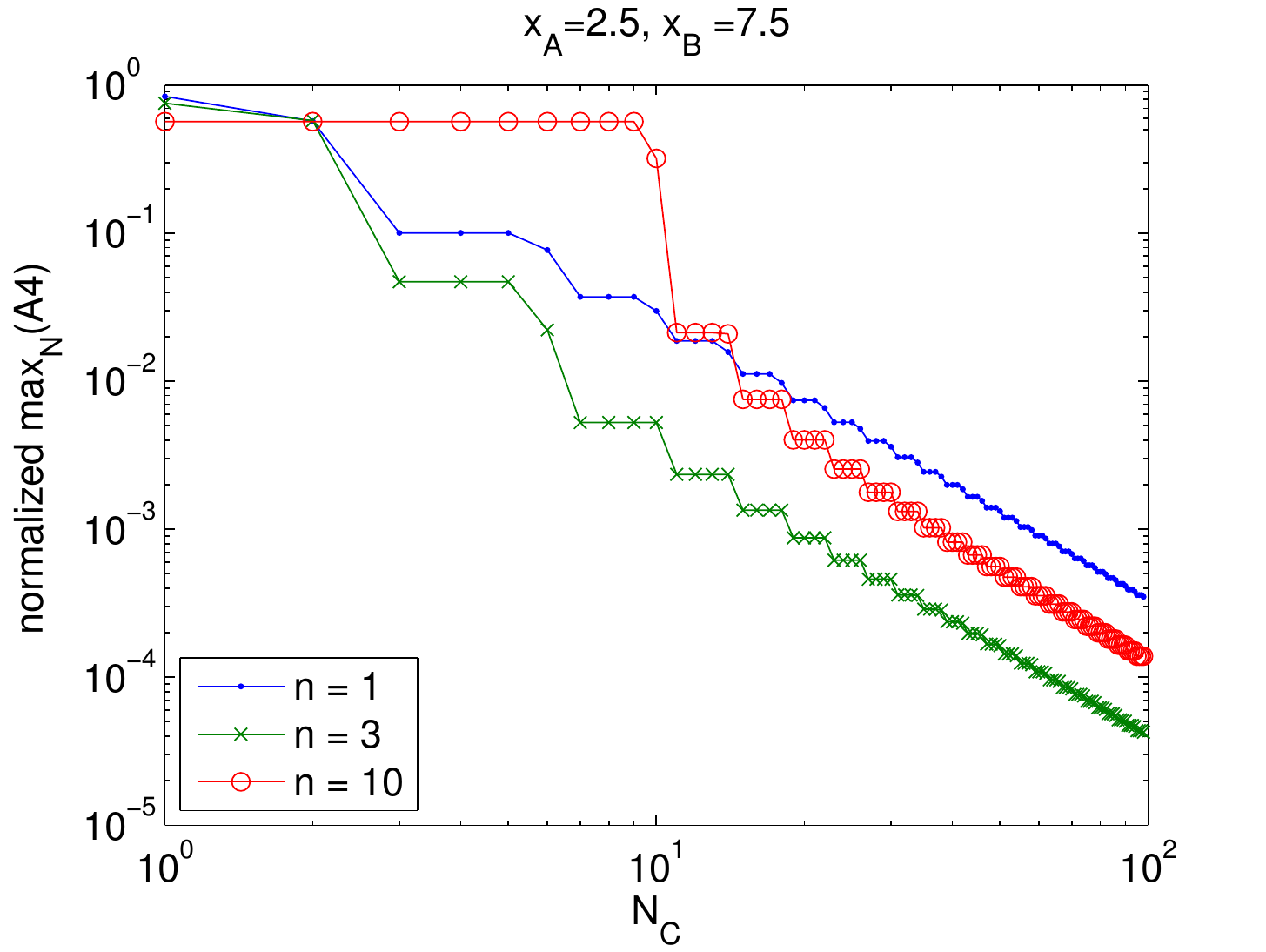}
  \caption{(Color online) The plot shows $\max_{N\geq N_C} |A_4(N)|$, the maximum value of the signalling term $A_4$ from \eqref{eq:A} on the light cone  \mbox{$|A_4(T=|x_\mathcal{A}-x_\mathcal{B}|)|$} obtained for any higher cutoff, i.e., for any number of modes $N\geq N_C$ larger than or equal to $N_C$.
For different $n$ the modes have been normalized as explained in the text.
The values of $|A_4|$ on the lightcone, i.e., for $T=|x_\mathcal{A}-x_\mathcal{B}|$ approach zero following a power law for large numbers of modes $N_C$. (All plotted quantities are dimensionless.)
}
  \label{fig:A4max}
\end{figure}

The most relevant figures are Fig. \ref{fig:A4ofT} and \ref{fig:A4max}. There we study the behaviour of $A_4(T)$ in the proximity of the light cone as the UV cutoff $N_C$ is increased.

In figure \ref{fig:A4ofT}, we see that for low values of $N_C$ the contribution $A_4$ is not vanishing for switching times $T<|x_\mathcal{A}-x_\mathcal{B}|$ shorter than the distance between the detectors. This would allow for superluminal signalling. Only with increasing $N_C$ the graph approaches the exact limit and causal behaviour is restored. We also observe in figure \ref{fig:A4ofT} that the values outside of the lightcone, i.e., for switching times $T<|x_\mathcal{A}-x_\mathcal{B}|$, grow with increasing switching times $T$.  Hence the lightcone where the switching time $T=|x_\mathcal{A}-x_\mathcal{B}|$ equals the distance between the detectors marks a critical case which we can use to quantify the violation of causality in a model with UV-cutoff: To avoid superluminal signalling the coefficients of the channel in \eqref{eq:chanstruct}, like $A$, have to vanish on 
the lightcone. As the value of $|A_4(T)|$ for a fixed cutoff $N_C$ is larger on the lightcone than further outside the lightcone, we can take $|A_4(T=|x_\mathcal{A}-x_\mathcal{B}|)|$ as a measure for the violation of causality.

In other words, given that the coefficients $A,B,C,D$ are smooth functions of time, their being zero outside the light cone implies that their value is also zero right on the light cone. If we are looking for an estimation of how big the   acausal error in signalling is  for a finite number of modes $N_C$, we can analyze the value of the contributions of Alice's detector initial state to Bob's detector right on the light cone, since this is the most conservative scenario.

Because of the oscillating behaviour that we observed in figure \ref{fig:A4ofNC} for large cutoffs, it is not convenient to directly compare values of $|A_4(T=|x_\mathcal{A}-x_\mathcal{B}|)|$ obtained for different values of $N_C$ to each other directly. Instead, in figure \ref{fig:A4max}, for a given cutoff $N_C$, we plot the maximum value obtained for $|A_4(T=|x_\mathcal{A}-x_\mathcal{B}|)|$ for any cutoff $N$ larger than or equal to $N_C$.

Notice that for all switching times $T$ the value of $A_4(T)$ depends also on the mode $n$ with which the detectors are resonant. In general, all contributions to the channel coefficients $A, B, C, D$ and $P$ in \eqref{eq:chanstruct} tend to be smaller for higher mode numbers $n$.  Therefore, in order to be able to compare the values for detectors being resonant with different modes $n$  on equal footing,  in figure \ref{fig:A4max}, we show the value of the $A_4$ term on the light cone divided by the respective value of $A_4$ for each $n$ at a time inside the lightcone. Hence we are computing the relative magnitude of the faster-than-light signalling signature as compared to the causal signal.
In particular, the values in figure \ref{fig:A4max} have been normalized by the respective value of $|A_4(T=\frac32 |x_A-x_B|, N_C=100)|$.

Interestingly, we observe that this value decays following a power law for cutoffs $N_C$ well above the resonance mode $n$.

The asymptotic power of this decay is  the same for different choices of the mode the detectors are tuned to be in resonance with. Similarly, the distance between the two detectors and their positioning inside the cavity do not change the slope of the decay, but only shift the asymptotic behaviour along the $y$-axis in a double-logarithmic plot as in figure \ref{fig:A4max}. This shift in the double-logaritmic plot corresponds to a multiplying factor in front of the functional relation between $N_C$ and $A_4$.

The power law decay can be traced back to the structure of $A_4$ as it is given in equation \eqref{eq:Aa} of the appendix. Inside the cavity the 2-point function of the field is given by a sum with a single contribution from each field mode. Hence the 4-point function which occurs in $A_4$ is given by a twofold sum with two summation variables running over all the field modes, because the 4-point function can be expressed in terms of 2-point functions in the usual way. For each term in this sum the time integrations of equation  \eqref{eq:Aa} lead to a polynomial in the summation variables (with the physical parameters and trigonometric functions as coefficients) which is divided by a common denominator. This denominator is itself a polynomial in the summation variables. Hence for higher cutoffs the asymptotic power law behaviour should emerge from the leading behaviour of the polynomial fraction for high values of the summation variables.


\subsection{Signalling using $\{\ket{+}, \ket{-}\}$ encoding}\label{sec:pmcoding}

We will discuss here how the use of $\ket{+}$ and $\ket{-}$ states can lead to a large improvement in the ability of Alice to signal Bob through the field as compared to preparing Alice's detector in the ground or excited states.

In \eqref{eq:chanstruct} we see that the off-diagonal elements of $\rho_{T,B}$ are given by products of the factors $C$ and $D$ (and their complex conjugates) with the off-diagonal elements $\gamma$ and $\gamma^*$ of $\rho_{A,0}$. So in general (i.e.,  unless $\rho_{\mathcal{A},0}$ is diagonal) the initial state of detector $\mathcal{A}$ has an influence on the final state of detector $\mathcal{B}$ at  second order in the coupling strength, because $C,D \sim \mathcal{O}(\lambda^2)$. However to make use of this effect, e.g., for signalling, the off-diagonal element of the input state $\gamma$ has to be large and on the recipients side, the off-diagonal elements of $\rho_{\mathcal{B},T}$ have to be measured.

As a simple example for this we look at the following protocol: Just as before,the system is assumed to start out in the state \eqref{eq:instate}, i.e., the field is prepared in the vacuum, detector $\mathcal{B}$ in its ground state and detector $\mathcal{A}$ in an arbitrary state
\begin{equation}
\rho_{\mathcal{A},0}= \begin{pmatrix}
\theta & \gamma \\ \gamma^* & \beta \end{pmatrix}.
\end{equation}
After the interaction has taken place between $t=0$ and $T=0$ a measurement on detector $\mathcal{B}$ is performed in the $\{\ket{+}, \ket{-}\}$ basis.

\begin{equation}\label{eq:pmstates}
\ket{\pm} = \frac{1}{\sqrt{2}} \left(\ket{g} \pm \ket{e} \right)
\end{equation}
we find that the projectors onto these two states are given by
\begin{equation}
P_\pm (t=0) = \proj{\pm}{\pm}= \frac12 \begin{pmatrix} 1 & \pm 1 \\ \pm 1 & 1 \end{pmatrix}  .
\end{equation}
Because we work in the interaction picture $P_\pm$ needs to be evolved with the corresponding free Hamiltonian.
\begin{multline}
P_\pm(t)= \exp(\ii H_0^{(\mathcal{D})} t) \, P_\pm(t=0) \, \exp (-\ii H_0^{(\mathcal{D})} t) \\ = \frac12 \begin{pmatrix}
1 & \pm e^{\ii \Omega_\mathcal{D} t} \\ \pm e^{-\ii \Omega_\mathcal{D} t}  & 1
\end{pmatrix}
\end{multline}
So for $t>T$ the probability to find detector $\mathcal{B}$, e.g.,  in the $\ket{+}$-state is given by:
\begin{equation}
\tr \left( P_+(t)  \rho_{T,B} \right) = \frac12 + \text{Re} \left( (\gamma C +\gamma^* D^*) \, e^{-\ii \Omega_\mathcal{B} t} \right)
\end{equation}
It is interesting to note, that this detection probability is completely independent of all other terms occuring in the general form of the channel. It is independent of $A$ and $B$ and hence from the diagonal elements of $\rho_{A,0}$ and the single detector excitation probability $P$ does not have any influence either.

\begin{figure}[htb] 
\includegraphics[width=0.48\textwidth]{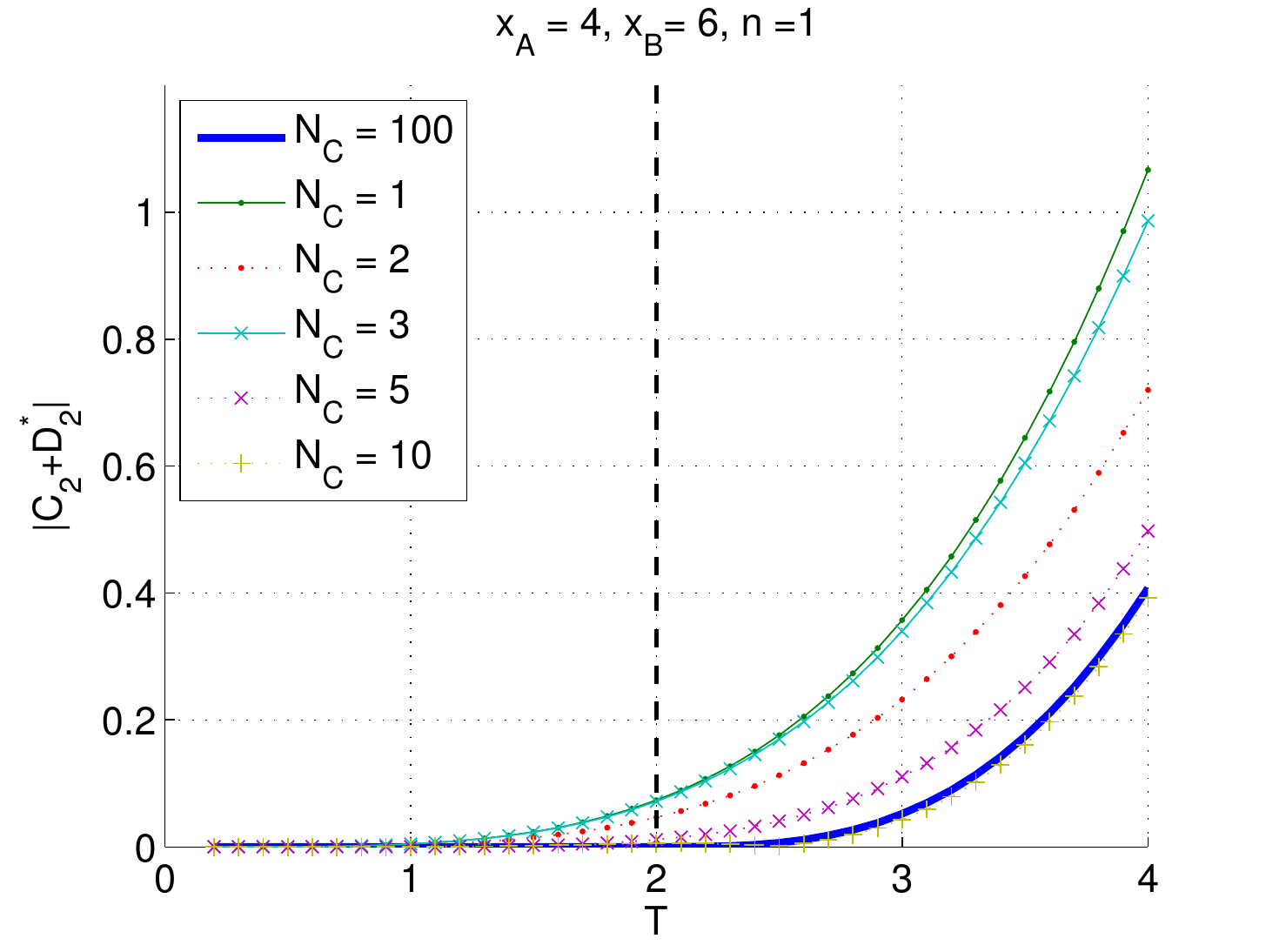}
  \caption{(Color online) Plot of $|C_2 + D_2^*|$ as defined in \eqref{eq:C}, \eqref{eq:D}, the lowest order contribution to the detection probability in \eqref{eq:pmprob}, for different cutoffs $N_C$. The detectors are located at a distance of $|x_A - x_B| =2$. The dashed line indicates the lightcone. The use of $\ket\pm$-states enhances signalling by two orders of magnitude in the coupling constant as compared to the use of energy eigenstates. (All plotted quantities are dimensionless.)
}
  \label{fig:C2D2}
\end{figure}

This could be of use for signalling between the two detectors: Say the detector $\mathcal{A}$ was intially prepared in the $\ket{+}$-state, for which $\gamma=\frac12$ or the $\ket{-}$-state, for which $\gamma=-\frac12$. Then for a given set of parameters, i.e., if $C$ and $D$ are known, the time point of the measurement on detector $\mathcal{B}$ can be chosen such, that the probability to find detector $\mathcal{B}$ in the $\ket{+}$-state is given by
\begin{equation}\label{eq:pmprob}
p\left(\mathcal{B}=\ket{+} | \mathcal{A}=\ket\pm\right) = \frac12 \pm \left| C + D^*\right|.
\end{equation}
Figure \ref{fig:C2D2} plots one example of the lowest order contributions to this probability.

The fact that $C,D\sim\mathcal{O}(\lambda^2)$ and that the detection probabilities \eqref{eq:pmprob} are independent of the other factors in the channel \eqref{eq:chanstruct} indicates that the protocol outlined above should allow for much enhanced signalling as compared to the use of eigenstates like in the Fermi problem. Although the preparation and detection of $\ket{\pm}$-states might be experimentally more difficult than the use of energy eigenstates. However for the energy eigenstates the leading order contributions to the detection probabilities are of order $\mathcal{O}(\lambda^4)$ and furthermore they compete with the single detector excitation probability $P\sim\mathcal{O}(\lambda^2)$ which also contributes to the diagonal elements of $\rho_{\mathcal{B},T}$.

\begin{figure}[htb] 
\includegraphics[width=0.48\textwidth]{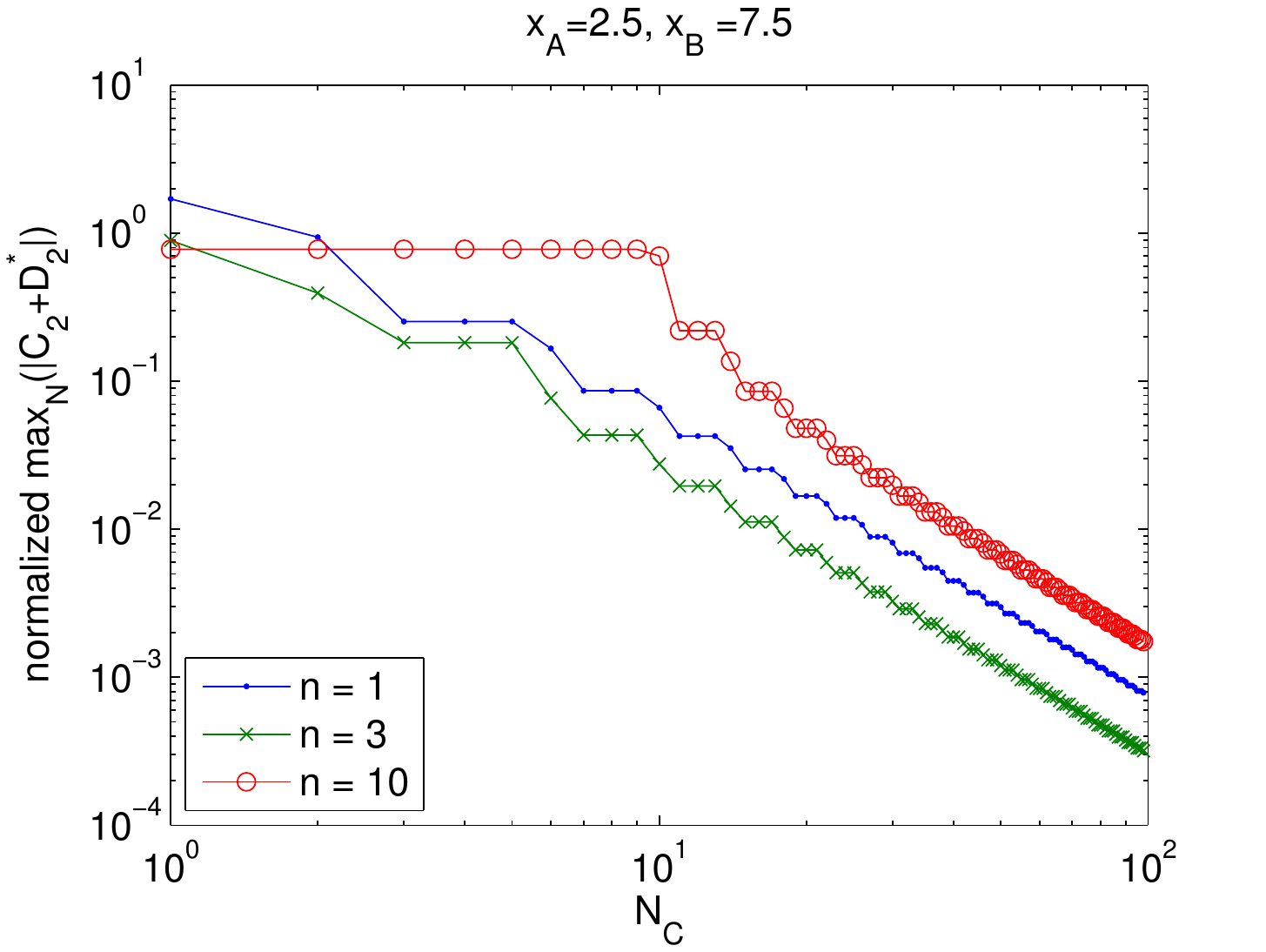}
  \caption{(Color online) The plot shows $\max_{N\geq N_C} |C_2 (N)+D^*_2 (N)|$ on the light cone, i.e., for $T=|x_A-x_B|=5$. Analogous to figure \ref{fig:A4max}, also here the values for different $n$ have been normalized in order to be able to compare them better. Again we observe a power law decay for high cutoffs $N_C$. (All plotted quantities are dimensionless.)}
  \label{fig:C2D2NCloglog}
\end{figure}

Figures \ref{fig:C2D2} and \ref{fig:C2D2NCloglog} show that the dependence of $|C_2 +D_2^*|$ on the size of the cutoff $N_C$ is similar to the behaviour obtained for the signalling term in the Fermi problem in the previous section. Figure \ref{fig:C2D2} illustrates the behaviour of close to the lightcone. If a too small number of modes are taken into account, the model is clearly inconsistent with causality but for higher and higher cutoffs the curve approaches the limit of full causality. In figure \ref{fig:C2D2NCloglog} we see that the errors again decay according to a power law, as already observed in the previous section.

\section{The channel in general scenarios and for arbitrary initial states of  $\mathcal{B}$}\label{sec:changen}
 After the previous analysis, one  may ask  how many of these results carry over to the continuum scenario and, additionally, what influence the initial state of detector $\mathcal{B}$ has as to which order in the coupling strength the signalling terms occur. So far we always assumed the initial state at $T=0$ to be of the form \eqref{eq:instate}: While detector $\mathcal{A}$ was free to be prepared in an arbitrary state, detector $\mathcal{B}$ always started out in its ground state $\rho_{\mathcal{B},0}=\proj{g}{g}$.

In this section we analyze the general structure of the quantum channel, analogously to \eqref{eq:chanstruct}, for arbitrary input states of detector $\mathcal{B}$. We find that the observations of the previous sections generalize. If $\mathcal{A}$ is prepared in an  eigenstate of the free Hamiltoinian, the influence of detector $\mathcal{A}$ on $\mathcal{B}$ is always suppressed by two orders of magnitude as compared to any other choice of pure input states.
The analysis and results of this section are as general as the introductory section \ref{sec:setting} and not restricted to fields inside a cavity, i.e., they equally apply to the light-matter interaction in the continuum scenario.

To allow arbitrary initial states for both detectors, we denote their initial density matrices as
\begin{equation}
\rho_{\mathcal{A},0} =\begin{pmatrix} \theta & \gamma \\ \gamma^* &  \beta \end{pmatrix},
\end{equation}
which is the same as in \eqref{eq:defnrhoA}, and 
\begin{equation}
\rho_{\mathcal{B},0} = \begin{pmatrix} \varphi & \delta \\ \delta^* & \kappa
\end{pmatrix}.
\end{equation}
Now the initial state of the entire system consisting of the two detectors and the field, which starts out in the vaccuum, reads
\begin{equation}
\rho_0 = \rho_{\mathcal{A},0}\otimes\rho_{\mathcal{B},0}\otimes \proj{0}{0}.
\end{equation}
The question is now how the final state 
\begin{equation}
\rho_{\mathcal{B},T}=\tr_{\mathcal{A},\mathcal{F}} [\rho_T] = \tr_{\mathcal{A},\mathcal{F}} \left[ U(T,0)\, \rho_T \,  U(T,0)^\dagger \right]
\end{equation}
of detector $\mathcal{B}$ depends on the elements of the inital density matrices. With the same kind of arguments as in section \ref{sec:setting}, that explained the structure of the channel in \eqref{eq:chanstruct}, and using the tracelessness and Hermiticity of the denisty matrix, we can deduce that the final state of detector $\mathcal{B}$ for general initial states of both detectors is of the form:

\begin{multline}\label{eq:chanstructgeneral}
\rho_{\mathcal{B},T}=\begin{pmatrix}
\varphi & \delta \\ \delta^* & \kappa
\end{pmatrix}+\begin{pmatrix}\kappa P + \varphi Q & \delta R + \delta^* S^* \\ \delta^* R^* +\delta S & -\kappa P -\varphi Q\end{pmatrix}\\
+ \gamma \begin{pmatrix} \delta I+\delta^* J & \kappa C+ \varphi G \\ \kappa D + \varphi H & -\delta I-\delta^* J  \end{pmatrix}\\
+ \gamma^* \begin{pmatrix}\delta J^* + \delta^* I^* & \kappa D^* + \varphi H^*\\ \kappa C^*+\varphi G^* &  -\delta J^* -\delta^* I^* \end{pmatrix}\\
+\theta \begin{pmatrix}
\kappa A+\varphi E & \delta K + \delta^* L^*\\ \delta L+\delta^* K^* & -\kappa A - \varphi E
\end{pmatrix}\\
+\beta \begin{pmatrix}
\kappa B+\varphi F & \delta M + \delta^* N^* \\ \delta N + \delta^* M^* & -\kappa B - \varphi F
\end{pmatrix}.
\end{multline}
Here $A,B,E,F,P,Q \in \mathbb{R}$ are real, whereas all other Latin letters stand for complex constants, that depend on the parameters, the geometry and the switching functions of the set-up. The constants $A,B,C,D,P$ are the ones which were already introduced in \eqref{eq:chanstruct}.

The constants multiplying $\gamma$ and $\gamma^*$, the off-diagonal elements of $\rho_{\mathcal{A},0}$, are all of order $\mathcal{O}(\lambda^2)$. Just as discussed for $C$ and $D$ earlier, all their perturbation expansion are of the form
\begin{equation}\label{eq:defn_ch_coefX}
X = \lambda_\mathcal{A} \lambda_\mathcal{B} X_2 +\mathcal{O}(\lambda^4), \text{ for } X= C,D,G,H,I,J.
\end{equation}
Whereas the terms that multiply the diagonal elements $\theta$ and $\beta$ of $\rho_{\mathcal{A},0}$ are of order $\mathcal{O}(\lambda^4)$.
\begin{equation}\label{eq:defn_ch_coefY}
Y = \lambda^2_\mathcal{A} \lambda^2_\mathcal{B} Y_4 + \mathcal{O}(\lambda^6), \text{ for } Y=A,B,E,F,K,L,M,N.
\end{equation}

 This means that signals that are encoded using energy eigenstates of the free detector are strongly suppressed as compared to any other choice of encoding.
This is because independently of the receiver's initial state, its final state is only influenced by the sender at order $\mathcal{O}(\lambda^4)$ with this encoding. That may be a problem for signal transmission because the quantum noise terms 
\begin{equation}\label{eq:defn_ch_coefZ}
Z = \lambda_\mathcal{B}^2 Z_2 +\mathcal{O}(\lambda_\mathcal{B}^4) , \text{ for } Z=P,Q,R,S
\end{equation}  
are of higher order in the coupling strength and might therefore overpower the signal. Hence, pure states with large off-diagonal elements, like the $\ket{\pm}$-states, should be a better choice for encoding, since they influence the receiver's final state at order $\mathcal{O}(\lambda^2)$. In this context, see also \cite{referee} where it was observed, studying correlation functions, that longitudinal correlations behave in a different way than transversal correlations. 


In Appendix \ref{app:calc} we calculate the leading order terms in the perturbative expansion of Bob's final state \eqref{eq:chanstructgeneral}, i.e., all the $\mathcal{O}(\lambda^2)$ contributions to the coefficients in \eqref{eq:defn_ch_coefX} and \eqref{eq:defn_ch_coefZ}. All second order contributions to the coefficients in \eqref{eq:defn_ch_coefX} are found to have the same absolute value (see \eqref{eq:G2a} to \eqref{eq:J2a}). Therefore the perturbative expansion of Bob's final state \eqref{eq:chanstructgeneral} simplifies to
\begin{align}\label{eq:pertexpr2}
\rho_{\mathcal{B},T} &= \begin{pmatrix}\varphi & \delta \\ \delta^* &  \kappa \end{pmatrix} \nonumber\\
&\quad +\lambda_\mathcal{B}^2\begin{pmatrix} \kappa P_2 + \varphi Q_2 & \delta R_2 + \delta^* S_2^* \\ \delta^* R_2^* + \delta S_2 & -\kappa P_2 - \varphi Q_2\end{pmatrix}\nonumber\\
& \quad +\lambda_\mathcal{A} \lambda_\mathcal{B} \left[ \gamma \begin{pmatrix} \delta D_2 + \delta^* C_2 & (\kappa-\varphi) C_2  \\ (\kappa -\varphi) D_2  & - \delta D_2 - \delta^* C_2\end{pmatrix} + \text{H.c.}\right]\nonumber\\
&\quad +\mathcal{O}(\lambda^4).
\end{align}

From this we see that to leading order in perturbation theory the strength of any signal sent from detector $\mathcal{A}$ to $\mathcal{B}$ is characterized just by the two integral expression for $C_2$ in \eqref{eq:C2a} and for $D_2$ in \eqref{eq:D2a}.

The initial state of the receiving detector $\mathcal{B}$ is not of as fundamental importance as the initial state of the sender, because in general its final state is always affected at order $\mathcal{O}(\lambda^2)$. (Except, possibly, for an unfortunate choice of initial states which depends on the actual values of the channel constants $A,...,N$.) However, it is interesting to see that one observation from the previous subsection generalizes: The contributions to the off-diagonal elements of $\rho_{\mathcal{B},T}$ are proportional to the diagonal elements $\varphi, \kappa$ of the initial state $\rho_{\mathcal{B},0}$ of the detector $\mathcal{B}$ and vice versa. This suggests that to detect the signals sent from detector $\mathcal{A}$, the final measurement on the state of detector $\mathcal{B}$ should be done in a basis which is orthogonal to the initial state.

This was the case in the protocol analyzed in section \ref{sec:pmcoding}, where $\mathcal{B}$ started out in the ground state but was finally measured in the $\ket{+}/\ket{-}$-basis. And from \eqref{eq:pertexpr2} we see, that if $\mathcal{B}$ starts out in a state $\rho_{\mathcal{B},0}$ such that $\varphi=\kappa$, i.e., an equal-weight superposition of $\ket e$ and $\ket g$ like the $\ket\pm$ states, then the off-diagonal terms of $\rho_{\mathcal{B},T}$ vanish. Hence the final measurement has to be done in the energy eigenbasis which is orthogonal to the initial state.

\section{Conclusions}

In practice, it is desirable for computational efficiency to truncate the mode expansion of a field inside a cavity, but the truncated model, in principle, suffers from noncausal signalling. We set out to determine how many modes need to be taken into account for the model to preserve causal signalling as a function of the desired smallness of the acausal error terms. We found that these error terms decay with the increase of the number of modes according to a universal power law, i.e., according to a power law that is independent of the initial state and the detector parameters. This means that suitable `few-mode approximations'  can be used to reliably model light-matter interactions in cavities even in the short-time regime.

Additionally, we found that amplitude modulation is sub-optimal in the short-time regime. In fact, it is the least efficient way to code with an orthogonal pair of states. Concretely, we found that if amplitude modulation is used, i.e., when energy eigenstates of the free detector are used to encode the signal, then the signal travels from one detector to the other only from the fourth order, $\mathcal{O}(\lambda^4)$, in perturbation theory. If instead superpositions of these states are used then the initial state of Alice influences Bob's final measurement outcome already to second order $\mathcal{O}(\lambda^2)$, which is the lowest order at which causal influence of $\mathcal{A}$ on $\mathcal{B}$ can possibly manifest itself. 
Since  the quantum noise also influences the detectors at order $\mathcal{O}(\lambda^2)$, a much better signal-to-(quantum)noise ratio is to be expected when coding with states other than $\ket{g},\ket{e}$.

As an illustrative example, we studied a protocol where $\ket\pm$-states where used to code the signal. We studied how the number of modes has to scale as a function of the desired accuracy with which causality is preserved and we found again that the scaling follows a certain power law. Finally we discussed that our results carry over to the case of Alice and Bob communicating through the exchange of quanta of a massless quantum field when the size of the cavity diverges, i.e., when one approaches a continuum of modes.   

To draw our final conclusions, let us recall the communication regime that we here considered. It is the regime where the coupling constants are small and, crucially, it is the regime of the earliest possible times when the signal could be picked up by Bob. This suggests an intuition for why amplitude modulation is sub-optimal in that regime: if Alice used $\ket{ e}$ states to signal, the emission of her signal would tend to be delayed because of the time that it can take the state $\ket{e}$ to decay and therefore to emit a field quantum that Bob could receive. What we found is that in the early-time regime, i.e., within roughly the half-life of Alice's excited state, any other coding is more efficient, as it relies not on the emission and absorption of a field quantum but on the immediately-starting overall change of the system that occurs when the interactions are switched on. 

\section{Outlook}

It should be very interesting to determine the change of the optimal coding scheme for communication through this channel 
when moving from the early-time regime to the late-time regime.  
Indeed, as figure \ref{fig:P2} shows, the noise term is bounded in time while one of the signalling terms when Alice uses amplitude coding, namely the term $A$, is growing in time. It grows even faster than the causal term, $C$, in the $\{\ket{+}, \ket{-}\}$ coding. Thus, even though $A$ is merely of fourth order in the coupling constant, given enough time it could in principle grow to the size of the term $C$ which is of second order in the coupling constant. The time scale at which this could happen is $T\approx \lambda^{-1}$, roughly the half lifetime of Alice's excited state. In this regime, when $T\lambda$ is no longer small,
the transition matrix elements of $-\ii H_I T$ are no longer small, i.e., the perturbation theory that we have used so far then breaks down. Here, it should be very interesting to apply the recently-developed non-perturbative methods for detectors that are harmonic oscillators \cite{UdWGauss,Fuenetesevolution} in order to extend our present study into the nonperturbative regime.

Further, our results suggest to re-investigate the intriguing generalized Huygens principle which is also called the lacuna effect \cite{mclenaghan1969explicit,mclenaghan1982huygens}. The lacuna effect is a phenomenon in the theory of differential equations. It implies that massless classical fields such as classical light only propagate on the light cone itself but not inside the light cone, in 3+1 dimensions, but, crucially, not in all dimensions. This means that in 3+1 dimensions, Bob's detector should have to be switched on exactly when Alice's signal passes through. A little later and he misses the chance to receive it, even though Bob is timelike to Alice's emission. 

With our new methods, and using pulsed switching of the detectors, it should be possible to model signalling using wave packets, to optimize the coding strategy and consequently to calculate the corresponding capacities explicitly. Once one can localize signal emission and absorption in time, it should then become possible to investigate the lacuna effect information theoretically, and perhaps gain new insight into the reasons for its mysterious dimension dependence. Notice that a connection between the Lacuna effect and the dimension-dependence of Hawking radiation has been proposed in \cite{ooguri1986spectrum}. There may well also exist a deep connection to studies of time-like vacuum entanglement, \cite{Olson2011,Olson2012,PastFutPRL}. 

The communication channel that we study here can be considered to be the basic prototype for any communication between simple quantum systems through the exchange of field quanta. It should be very interesting, therefore, to study this channel in various key circumstances, for example, where Alice and/or Bob are accelerated, where there is curvature, a horizon or a potential, or in the presence of a dissipative environment. 

It may also be of interest to extend this study to a scenario of quantum broadcasting, with many potential senders and receivers. Indeed, ultimately, thinking speculatively, any interaction between systems through the exchange of intermediate particles, including even the interactions among elementary particles, may become describable information theoretically, namely in terms of the flow of quantum information through those interactions.  This may be a desirable route to go because the unit of information, the qubit, is more robust than, for example, the units of mass or distance, which tend to loose their operational meaning in sufficiently extreme circumstances, such as in quantum gravity phenomena.


\section{Acknowledgements} The authors gratefully acknowledge support through the Ontario Trillium Scholarship and NSERC's Banting, Discovery and Canada Research Chairs Programs. AK acknowledges the kind hospitality during his sabbatical stay at the Centre for Quantum Computation \& Communication Technology at the University of Queensland.

\appendix

\begin{widetext}

\section{Integral form of the channel coefficients}\label{AppA}

In this appendix we present the integral form of the coefficients $A,B$ and $P$ defined in equations \eqref{eq:P},\eqref{eq:A} and \eqref{eq:B}. The lowest order integrals for the coefficients $C$ and $D$ are given in \eqref{eq:C2a} and \eqref{eq:D2a}.

\begin{align}
A &=\lambda_\mathcal{A}^2 \lambda_\mathcal{B}^2 \, A_4+\mathcal{O}(\lambda^6)\nonumber\\
&= \lambda_\mathcal{A}^2 \lambda_\mathcal{B}^2 \integral{t_1}{0}T \integral{t_2}{0}{t_1} \integral{s_1}{0}T \integral{s_2}{0}{s_1} \nonumber\\
&\qquad \times \left[ \chi_\mathcal{A}(t_1) \chi_\mathcal{A}(s_2) \chi_\mathcal{B}(t_2) \chi_\mathcal{B}(s_1) \left( e^{\ii \Omega_\mathcal{B} (t_2-s_1) -\ii \Omega_\mathcal{A}(t_1-s_2)} \fourpoint{A}{s_2}B{s_1}A{t_1}B{t_2} + \text{H.c.} \right)  \right. \nonumber \\
& \qquad\quad + \left. \chi_\mathcal{A}(t_1) \chi_\mathcal{A}(s_1) \chi_\mathcal{B}(t_2) \chi_\mathcal{B}(s_2) \left( e^{\ii \Omega_\mathcal{B} (t_2-s_2) -\ii \Omega_\mathcal{A} (t_1-s_1)} \fourpoint{B}{s_2}A{s_1}A{t_1}B{t_2}  \right)  \right. \nonumber\\
 &\qquad\quad + \left. \chi_\mathcal{A}(t_2) \chi_\mathcal{A}(s_2) \chi_\mathcal{B}(t_1) \chi_\mathcal{B}(s_1) \left( e^{\ii \Omega_\mathcal{B} (t_1-s_1) -\ii \Omega_\mathcal{A} (t_2-s_2)} \fourpoint{A}{s_2}B{s_1}B{t_1}A{t_2}  \right)  \right] \nonumber \\ 
&\quad - \lambda_\mathcal{A}^2 \lambda_\mathcal{B}^2 \integral{s}{0}T \integral{t_1}{0}T \integral{t_2}{0}{t_1} \integral{t_3}{0}{t_2} \nonumber\\
&\qquad \times \left[ \chi_\mathcal{A}(t_1) \chi_\mathcal{A}(t_2) \chi_\mathcal{B}(t_3) \chi_\mathcal{B}(s) \left( e^{\ii \Omega_\mathcal{B} (t_3-s) +\ii \Omega_\mathcal{A} (t_1-t_2)} \fourpoint{B}s{A}{t_1}A{t_2}B{t_3} + \text{H.c.} \right) \right. \nonumber\\
&\qquad \quad + \left. \chi_\mathcal{A}(t_1) \chi_\mathcal{A}(t_3) \chi_\mathcal{B}(t_2) \chi_\mathcal{B}(s) \left( e^{\ii \Omega_\mathcal{B} (t_2-s) +\ii \Omega_\mathcal{A} (t_1-t_3)} \fourpoint{B}s{A}{t_1}B{t_2}A{t_3} + \text{H.c.} \right) \right. \nonumber\\
&\qquad\quad + \left. \chi_\mathcal{A}(t_2) \chi_\mathcal{A}(t_3) \chi_\mathcal{B}(t_1) \chi_\mathcal{B}(s) \left( e^{\ii \Omega_\mathcal{B} (t_1-s) +\ii \Omega_\mathcal{A} (t_2-t_3)} \fourpoint{B}s{B}{t_1}A{t_2}A{t_3} + \text{H.c.} \right) \right] \nonumber \\ 
&\quad +\mathcal{O}(\lambda^6)\label{eq:Aa}\\
B&= \lambda_\mathcal{A}^2 \lambda_\mathcal{B}^2 B_4 +\mathcal{O}(\lambda^6) = A(-\Omega_\mathcal{A},\Omega_\mathcal{B})\label{eq:Ba} \\
P&= \lambda_\mathcal{B}^2 \,  P_2 + \lambda_\mathcal{B}^4 \, P_4 +\mathcal{O}(\lambda_\mathcal{B}^6)\nonumber\\
&=\lambda_\mathcal{B}^2 \integral{t_1}{0}T \integral{t_2}{0}T \, \chi_\mathcal{B}(t_1) \chi_\mathcal{B}(t_2) \, e^{\ii \, \Omega_\mathcal{B} (t_1-t_2)} \left< \phi(x_\mathcal{B}(t_2)) \phi( x_\mathcal{B}(t_1)) \right>\nonumber\\
&\quad-\lambda_\mathcal{B}^4 \integral{s}{0}T \integral{t_1}{0}T \integral{t_2}{0}{t_1}\integral{t_3}{0}{t_2} \nonumber\\
&\qquad\quad \times \left[ \chi_\mathcal{B}(t_1) \chi_\mathcal{B}(t_2) \chi_\mathcal{B}(t_3) \chi_\mathcal{B}(s)
\left( e^{\ii \Omega_\mathcal{B} (t_1-t_2+t_3-s)} \fourpoint{B}s{B}{t_1}B{t_2}B{t_3} + \text{H.c.} \right)\right] \nonumber\\
&\quad +\mathcal{O}(\lambda_\mathcal{B}^6)  \label{eq:Pa}
\end{align}

Notice that in comparison to \cite{mathieuachim1} the terms for $A$ and $B$ have a different integral structure, because we derived them using the Dyson expansion as in \eqref{eq:rhobout}. In this form not all the integral boundaries are dependent on each other, which should be an advantage for numerical evaluations.
Also, as mentioned in section \ref{sec:setting}, we can obtain two different expressions for the contributions to $P, A$ and $B$ from \eqref{eq:rhobout}, one of which comes as the coefficient of $\proj{e_{\mathcal{B}}}{e_{\mathcal{B}}}$ whereas the other the comes with $\proj{g_{\mathcal{B}}}{g_{\mathcal{B}}}$. These two forms have different integral structures but are, of course, equivalent. 

\section{Integral form of the channel coefficients}\label{app:calc}
In this appendix we give a detailed calculation of the  $\mathcal{O}(\lambda^2)$ contributions to \eqref{eq:chanstructgeneral}. These are the leading order contributions to Bob's final density matrix in the most general case where Alice and Bob are allowed to start out in arbitrary initial states while the field still starts out in the vacuum state.

The interaction Hamiltonian in the interaction picture for two Unruh-DeWitt detectors coupled to the Klein-Gordon field is the sum of two single detector interaction Hamiltonians $\intH{A}$ and $\intH{B}$:
\begin{align}
H_I (t) &= \intH{A}+\intH{B}
= \sum_{d=\mathcal{A},\mathcal{B}} \lambda_\mathcal{D} \,\chi_\mathcal{D}(t)\, \mu_\mathcal{D}(t)\, \phi(x_\mathcal{D}(t))
= \sum_{\mathcal{D}=\mathcal{A},\mathcal{B}}\,\mud{D}(t) \,\phi(x_\mathcal{D}(t)).
\end{align}
Here we introduced the shorthand notation ${\mud{D}} (t)=\lambda_\mathcal{D} \chi_\mathcal{D}(t) \mu_\mathcal{D}(t)$.
As in section \ref{sec:changen} we denote the initial state of the system by
\begin{align}
\rho_0 &= \underbrace{\rho_{\mathcal{A},0} \otimes \rho_{\mathcal{B},0} }_{\rho_{\mathcal{AB}}} \otimes \ket{0} \bra{0}
=\begin{pmatrix}
\theta & \gamma \\ \gamma^* & \beta
\end{pmatrix}
\otimes
\begin{pmatrix}
\varphi & \delta \\ \delta^* & \kappa
\end{pmatrix}
\otimes
\ket{0}\bra{0}.
\end{align}

From equation \eqref{eq:rho2} the second order perturbative corrections $\rho_T^{(2)}$ to the total system's final state $\rho_T$ are given by
\begin{align}\label{eq:apprho2}
\rho_T^{(2)} &= U^{(2)} \rho_0 + \rho_0 {U^{(2)}}^\dagger + U^{(1)} \rho_0 {U^{(1)}}^\dagger\nonumber\\&
=-\left( \integral{t_1}{0}{T} \integral{t_2}0{t_1} \, H_I(t_1) \, H_I(t_2) \right) \rho_0 -\rho_0 \left(\integral{t_1}0T \integral{t_2}0{t_1} \, H_I(t_2)\, H_I(t_1)\right)\nonumber\\&\quad
 +\left(\integral{t_1}0T H_I(t_1)\right) \rho_0 \left(\integral{t_2}0T H_I(t_2)\right)\nonumber\\
&=-\integral{t_1}0T\integral{t_2}0{t_1} \left.\left. \Big[  \Big( \intH{A}(t_1) \intH A(t_2) +\intH A(t_1) \intH B(t_2) + \intH B (t_1) \intH A(t_2) +\intH B(t_1) \intH B (t_2)\right) \rho_0 \right.\nonumber\\
&\phantom{=-\integral{t_1}0T\integral{t_2}0{t_1} \Big[ } \quad + \left. \rho_0 \left( \intH A(t_2) \intH A(t_1) + \intH A (t_2) \intH B (t_1) + \intH B (t_2) \intH A (t_1) + \intH B (t_2) \intH B (t_1)\Big)\Big] \right. \right.\nonumber\\
&\quad +\integral{t_1}0T \integral{t_2}0T \left.\Big[ \intH A(t_1) \rho_0 \intH A (t_2) + \intH A(t_1) \rho_0 \intH B(t_2) + \intH B(t_1) \rho_0 \intH A(t_2) + \intH B(t_1) \rho_0 \intH B(t_2)\Big]\right. .
\end{align}
Bob's final density matrix $\rho_{\mathcal{B},T}$ is obtained by taking the partial trace over the field's and Alice's subspace (see \eqref{eq:channel}). Taking the partial trace of \eqref{eq:apprho2} over the field first leaves us with a two-point function in each term:
\begin{align}
\tr_\mathcal{F} \rho^{(2)} &= - \integral{t_1}0T \integral{t_2}0{t_1} \left. \Big[ \left.\Big( \mud A (t_1) \mud A(t_2) \twopoint A{t_1}A{t_2} + \mud A(t_1) \mud B(t_2) \twopoint A{t_1}B{t_2} \right. \right.\nonumber\\
&\phantom{= - \integral{t_1}0T \integral{t_2}0{t_1} \Big[ }\quad \left. \left. + \mud B(t_1) \mud A(t_2) \twopoint B{t_1}A{t_2} + \mud B(t_1) \mud B(t_2) \twopoint B{t_1}B{t_2}\Big)\right. \rho_\mathcal{AB} \right. \nonumber\\
&\phantom{= - \integral{t_1}0T \integral{t_2}0{t_1} } +\left. \rho_\mathcal{AB} \left. \Big( \mud A(t_2) \mud A(t_1) \twopoint A{t_2}A{t_1} + \mud A(t_2) \mud B(t_1) \twopoint A{t_2}B{t_1} \right.\right.\nonumber\\
&\phantom{=- \integral{t_1}0T \integral{t_2}0{t_1} + \rho_\mathcal{A} }\quad \left.\left. +\mud B(t_2) \mud A(t_1) \twopoint B{t_2}A{t_1} + \mud B (t_2) \mud B(t_1) \twopoint B{t_2}B{t_1} \Big) \Big]\right.\right. \nonumber\\
&\quad + \integral{t_1}0T\integral{t_2}0T \left. \Big[ \mud A(t_1) \rho_\mathcal{AB} \mud B(t_2) \twopoint A{t_2}A{t_1}+ \mud A(t_1) \rho_\mathcal{AB} \mud B(t_2) \twopoint B{t_2}A{t_1} \right. \nonumber\\
&\phantom{\quad + \integral{t_1}0T\integral{t_2}0T  }\quad \left. + \mud B(t_1) \rho_\mathcal{AB} \mud A(t_2) \twopoint A{t_2}B{t_1} + \mud B(t_1) \rho_\mathcal{AB} \mud B(t_2) \twopoint B{t_2}B{t_1}  \Big].\right.
\end{align}
Next we take the partial trace over Alice's detector. At this point the terms that describe interactions only between Alice and the field, but leave Bob's detector unaffected, drop out. This is because, as we showed in section \ref{sec:setting}, the second order contributions to a system consisting of only one detector and the field have vanishing trace. For the terms that contain one factor of $\mud A(t)$ the partial trace gives a scalar factor which reads
\begin{align}
\Gamma_\mathcal{A}(t):= \tr \mud A(t) \rho_\mathcal{A} = \tr  \rho_\mathcal{A} \mud A(t) = \lambda_\mathcal{A} \chi_\mathcal{A}(t) \left( \gamma e^{-i \Omega_\mathcal{A} t} + \gamma^* e^{i \Omega_\mathcal{A} t} \right).
\end{align}
With this definition we obtain
\begin{align}
\rho^{(2)}_{\mathcal{B},T} &= \tr_{\mathcal{A}}\tr_\mathcal{F} \rho^{(2)} \nonumber \\ 
&= -\integral{t_1}0T \integral{t_2}0{t_1} \left. \Big[ \Gamma_\mathcal{A}(t_1) \twopoint A{t_1}B{t_2} \mud B(t_2) \rho_{\mathcal{B},0} +\Gamma_\mathcal{A}(t_2) \twopoint B{t_1}A{t_2} \mud B(t_1) \rho_{\mathcal{B},0} \right. \nonumber\\
&\phantom{= -\integral{t_1}0T \integral{t_2}0{t_1}}\quad +\Gamma_\mathcal{A}(t_2) \twopoint A{t_2}B{t_1} \rho_{\mathcal{B},0} \mud B(t_1) + \Gamma_\mathcal{A} (t_1) \twopoint B{t_2}A{t_1} \rho_{\mathcal{B},0} \mud B(t_2) \nonumber\\
&\phantom{= -\integral{t_1}0T \integral{t_2}0{t_1}}\quad \left. + \twopoint B{t_1}B{t_2} \mud B(t_1) \mud B(t_2) \rho_{\mathcal{B},0}+ \twopoint B{t_2}B{t_1} \rho_{\mathcal{B},0} \mud B(t_2) \mud B(t_1) \Big] \right. \nonumber\\
&\quad + \integral{t_1}0T \integral{t_2}0T \left. \Big[ \Gamma_\mathcal{A} (t_1) \twopoint B{t_2}A{t_1} \rho_{\mathcal{B},0} \mud B(t_2) + \Gamma_\mathcal{A} (t_2) \twopoint A{t_2}B{t_1} \mud B(t_1) \rho_{\mathcal{B},0} \right.\nonumber\\
&\phantom{\quad + \integral{t_1}0T \integral{t_2}0T}\quad \left. + \twopoint B{t_2}B{t_1} \mud B(t_1) \rho_{\mathcal{B},0} \mud B(t_2) \Big]\right. .
\end{align}
Inserting the definitions of $\Gamma_\mathcal{A}$ and $\mud{B}$ and switching to matrix notation for Bob's density matrix gives
\begin{align}\label{eq:apprho2exp}
\rho^{(2)}_{\mathcal{B},T} &=-\integral{t_1}0T\integral{t_2}0{t_1} \left. \Big[ \lambda_\mathcal{A} \lambda_\mathcal{B} \left(  \chi_\mathcal{A}(t_1) \chi_\mathcal{B}(t_2) \left(\gamma e^{-i \Omega_\mathcal{A} t_1} + \gamma^* e^{i \Omega_\mathcal{A} t_1}\right) \twopoint A{t_1}B{t_2} \begin{pmatrix} \delta^* e^{i\Omega_\mathcal{B} t_2} & \kappa e^{i \Omega_\mathcal{B} t_2} \\ \varphi e^{-i \Omega_\mathcal{B} t_2} & \delta e^{-i \Omega_\mathcal{B} t_2} \end{pmatrix} \right. \right. \nonumber\\
&\phantom{=-\integral{t_1}0T\integral{t_2}0{t_1} } \phantom{\lambda_\mathcal{A} \lambda_\mathcal{B}} \quad \, \left. \left.+ \chi_\mathcal{A}(t_2) \chi_\mathcal{B}(t_1) \left(\gamma e^{-i \Omega_\mathcal{A} t_2} + \gamma^* e^{i \Omega_\mathcal{A} t_2}\right) \twopoint B{t_1}A{t_2} \begin{pmatrix} \delta^* e^{i\Omega_\mathcal{B} t_1} & \kappa e^{i \Omega_\mathcal{B} t_1} \\ \varphi e^{-i \Omega_\mathcal{B} t_1} & \delta e^{-i \Omega_\mathcal{B} t_1} \end{pmatrix}\right. \right. \nonumber\\
&\phantom{=-\integral{t_1}0T\integral{t_2}0{t_1} } \phantom{\lambda_\mathcal{A} \lambda_\mathcal{B}} \quad \, \left. \left.+ \chi_\mathcal{A}(t_2) \chi_\mathcal{B}(t_1) \left(\gamma e^{-i \Omega_\mathcal{A} t_2} + \gamma^* e^{i \Omega_\mathcal{A} t_2}\right) \twopoint A{t_2}B{t_1} \begin{pmatrix} \delta e^{-i\Omega_\mathcal{B} t_1} & \varphi e^{i \Omega_\mathcal{B} t_1} \\ \kappa e^{-i \Omega_\mathcal{B} t_1} & \delta^* e^{i \Omega_\mathcal{B} t_1} \end{pmatrix}\right. \right. \nonumber\\
&\phantom{=-\integral{t_1}0T\integral{t_2}0{t_1} } \phantom{\lambda_\mathcal{A} \lambda_\mathcal{B}} \quad \, \left. \left.+ \chi_\mathcal{A}(t_1) \chi_\mathcal{B}(t_2) \left(\gamma e^{-i \Omega_\mathcal{A} t_1} + \gamma^* e^{i \Omega_\mathcal{A} t_1}\right) \twopoint B{t_2}A{t_1} \begin{pmatrix} \delta e^{-i\Omega_\mathcal{B} t_2} & \varphi e^{i \Omega_\mathcal{B} t_2} \\ \kappa e^{-i \Omega_\mathcal{B} t_2} & \delta^* e^{i \Omega_\mathcal{B} t_2} \end{pmatrix} \right) \right. \nonumber\\
&\phantom{=-\integral{t_1}0T\integral{t_2}0{t_1}} \quad \left. +\lambda_\mathcal{B}^2 \, \chi_\mathcal{B}(t_1) \chi_\mathcal{B}(t_2) \left( \twopoint B{t_1}B{t_2} \begin{pmatrix} \varphi e^{i \Omega_\mathcal{B} (t_1 - t_2)} & \delta e^{i \Omega_\mathcal{B} (t_1-t_2)} \\ \delta^* e^{-i \Omega_\mathcal{B} (t_1 - t_2)} & \kappa e^{-i \Omega_\mathcal{B}(t_1 - t_2)}\end{pmatrix} \right. \right. \nonumber\\
&\phantom{=-\integral{t_1}0T\integral{t_2}0{t_1} \quad +\lambda_\mathcal{B}^2\, \chi_\mathcal{B}(t_1) \chi_\mathcal{B}(t_2) }  \quad \left. \left. + \twopoint B{t_2}B{t_1} \begin{pmatrix} \varphi e^{-i \Omega_\mathcal{B} (t_1 - t_2)} & \delta e^{i \Omega_\mathcal{B} (t_1-t_2)} \\ \delta^* e^{-i \Omega_\mathcal{B} (t_1 - t_2)} & \kappa e^{i \Omega_\mathcal{B}(t_1 - t_2)}\end{pmatrix} \right) \right] \nonumber\\
&\quad + \integral{t_1}0T\integral{t_2}0T \left[ \lambda_\mathcal{A} \lambda_\mathcal{B} \left( \chi_\mathcal{A}(t_1) \chi_\mathcal{B}(t_2) \left(\gamma e^{-i \Omega_\mathcal{A} t_1} + \gamma^* e^{i \Omega_\mathcal{A} t_1}\right) \twopoint B{t_2}A{t_1} \begin{pmatrix} \delta e^{-i \Omega_\mathcal{B} t_2} & \varphi e^{i \Omega_\mathcal{B} t_2} \\ \kappa e^{-i \Omega_\mathcal{B} t_2} & \delta^* e^{i \Omega_\mathcal{B} t_2} \end{pmatrix} \right. \right. \nonumber\\
&\phantom{\quad + \integral{t_1}0T\integral{t_2}0T  \lambda_\mathcal{A} \lambda_\mathcal{B} } \quad \left. \left. \,+ \chi_\mathcal{A}(t_2) \chi_\mathcal{B}(t_1) \left(\gamma e^{-i \Omega_\mathcal{A} t_2} + \gamma^* e^{i \Omega_\mathcal{A} t_2}\right) \twopoint A{t_2}B{t_1} \begin{pmatrix} \delta^* e^{i \Omega_\mathcal{B} t_1} & \kappa e^{i \Omega_\mathcal{B} t_1} \\ \varphi e^{-i \Omega_\mathcal{B} t_1} & \delta e^{-i \Omega_\mathcal{B} t_1} \end{pmatrix} \right) \right. \nonumber\\
&\phantom{\quad + \integral{t_1}0T\integral{t_2}0T} \quad \left. +\lambda_\mathcal{B}^2 \, \chi_\mathcal{B}(t_1) \chi_\mathcal{B}(t_2)\twopoint B{t_2}B{t_1} \begin{pmatrix} \kappa e^{i \Omega_\mathcal{B} (t_1-t_2)} & \delta^* e^{i \Omega_\mathcal{B} (t_1+t_2)} \\ \delta e^{-i \Omega_\mathcal{B} (t_1 + t_2)} & \varphi e^{-i \Omega_\mathcal{B} (t_1 - t_2)}\end{pmatrix}\right].
\end{align}
From this expression we can read off the second order contributions to the channel coefficients by comparison to \eqref{eq:chanstructgeneral}.
For the noise terms $P,Q,R,S$ from \eqref{eq:defn_ch_coefZ} the lowest order contributions are found to be:
\begin{align}
P_2&= \integral{t_1}0T\integral{t_2}0T\, \chi_\mathcal{B}(t_1) \chi_\mathcal{B}(t_2)  e^{i \Omega_\mathcal{B} (t_1 - t_2)} \twopoint B{t_2}B{t_1}\\
Q_2 &= -\integral{t_1}0T\integral{t_2}0T\,  \chi_\mathcal{B}(t_1) \chi_\mathcal{B}(t_2)  e^{-i \Omega_\mathcal{B} (t_1 - t_2)} \twopoint B{t_2}B{t_1}\\
R_2 &= -\integral{t_1}0T \integral{t_2}0{t_1} \, \chi_\mathcal{B}(t_1) \chi_\mathcal{B}(t_2) e^{i \Omega_\mathcal{B} (t_1 -t_2)} \left( \twopoint B{t_1}B{t_2} + \twopoint B{t_2}B{t_1} \right)\\
S_2 &= \integral{t_1}0T \integral{t_2}0T \, \chi_\mathcal{B}(t_1) \chi_\mathcal{B}(t_2)  e^{-i \Omega_\mathcal{B} (t_1+t_2)} \twopoint B{t_2}B{t_1}.
\end{align}

The lowest order signalling terms  from \eqref{eq:defn_ch_coefX} contain a sum of different integral terms from \eqref{eq:apprho2exp}. These integrals can be combined, e.g., for $C$ we have
\begin{align}\label{eq:C2a}
C_2 &= -\integral{t_1}0T\integral{t_2}0{t_1} \left( \chi_\mathcal{A}(t_1) \chi_\mathcal{B} (t_2) e^{i (\Omega_\mathcal{B} t_2 -\Omega_\mathcal{A} t_1)} \twopoint A{t_1}B{t_2}  \right. \nonumber\\
&\phantom{= -\integral{t_1}0T\integral{t_2}0{t_1}}\quad
\left. + \chi_\mathcal{A}(t_2) \chi_\mathcal{B} (t_1) e^{i (\Omega_\mathcal{B} t_1 -\Omega_\mathcal{A} t_2)} \twopoint B{t_1}A{t_2} \right) \nonumber\\
&\quad + \integral{t_1}0T \integral{t_2}0T \, \chi_\mathcal{A}(t_2) \chi_\mathcal{B} (t_1) e^{i(\Omega_\mathcal{B} t_1 - \Omega_\mathcal{A} t_2)} \twopoint A{t_2}B{t_1}\nonumber\\
&= \integral{t_1}0T \integral{t_2} 0 {t_1}\, \chi_\mathcal{A} (t_2) \chi_\mathcal{B} (t_1) e^{i (\Omega_\mathcal{B} t_1 - \Omega_\mathcal{A} t_2)} \comm{\phi (x_\mathcal{A}(t_2))}{\phi( x_\mathcal{B} (t_1))} 
\end{align}
and for $D$ we find
\begin{align}\label{eq:D2a}
D_2&= \integral{t_1}0T \integral{t_2} 0 {t_1}\, \chi_\mathcal{A} (t_2) \chi_\mathcal{B} (t_1) e^{-i (\Omega_\mathcal{B} t_1 +\Omega_\mathcal{A} t_2)} \comm{\phi( x_\mathcal{B} (t_1))}{\phi (x_\mathcal{A}(t_2))}.
\end{align}
Here, a sign error in \cite{mathieuachim1} in the exponent of \eq{eq:D2a} has been corrected.
The remaining lowest order contributions can be expressed in terms of $C_2$ and $D_2$:
\begin{align}
G_2&= - C_2 \label{eq:G2a}\\
H_2&= - D_2\\
I_2&=D_2\\
J_2&=C_2.\label{eq:J2a}
\end{align}



\end{widetext}

\bibliography{cavity_refs}

\end{document}